\documentclass[sigconf]{acmart}

\newcommand{\name}{\textsc{CrossLink}\xspace}

\newif\ifshowrev
\showrevfalse
\newcommand{\rev}[1]{\ifshowrev\textcolor{blue}{#1}\else#1\fi}

\usepackage{tcolorbox}
\tcbuselibrary{skins,breakable}
\usepackage{booktabs, graphicx}
\usepackage{listings}
\usepackage{xspace}
\usepackage{amsmath}

\usepackage{amsfonts}
\usepackage{amsthm}
\newtheorem{theorem}{Theorem}

\usepackage{balance}
\usepackage{xcolor}
\usepackage{placeins}

\usepackage{amsthm} 

\usepackage{enumerate} 
\usepackage{url}
\usepackage{hyperref}

\usepackage{tikz}
\usetikzlibrary{positioning, matrix, calc, fit}

\tikzset{
  rulelabel/.style={circle,draw=black,fill=red!20,inner sep=1pt}
}

\usepackage{booktabs}
\usepackage{array}
\usepackage{adjustbox}
\usepackage{caption}
\usepackage{subcaption}
\usepackage[capitalize]{cleveref}
\usepackage{graphicx}
\newtheorem{definition}{Definition}

\usepackage[linesnumbered,ruled,vlined]{algorithm2e}
\usepackage{algpseudocode, float}
\newcounter{algsubcounter}

\usepackage{tabularx}

\usepackage{array}
\usepackage{comment}

\usepackage{pifont}
\newcommand{\cmark}{\ding{51}} 
\newcommand{\xmark}{\ding{55}}

\usepackage{booktabs}
\usepackage{tabularx}
\usepackage{xurl}
\newcommand{\macid}[1]{\begingroup\urlstyle{tt}\path{#1}\endgroup}

\newcommand{\code}[1]{\textsf{#1}}
\newcommand{\protid}[1]{\ensuremath{\code{ID}_{#1}}\xspace}
\newcommand{\protidname}[2]{\ensuremath{\code{ID}_{#1}^{#2}}\xspace}
\newcommand{\linksto}{\leftrightsquigarrow}

\newcommand{\InterMap}[2]{\code{InterMap}_{{#1},{#2}}}
\newcommand{\did}[1]{d_{\code{ID}_{#1}}}
\newcommand{\pid}{\ensuremath{p_{\code{ID}}}}
\newcommand{\tid}{\ensuremath{t_{\code{ID}}}}

\newcommand{\data}{D}
\newcommand{\identifiers}{\code{ID}s}

\newcommand{\id}{\code{ID}}
\newcommand{\setidin}[1]{\ensuremath{S_{\code{#1}}}}

\newcommand{\IntraMap}[1]{\code{IntraMap}_{{#1}}}
\newcommand{\dhat}{\ensuremath{\widehat{d}}}

\newif\iffullversion
\fullversionfalse
\ifdefined\FULLVERSION \fullversiontrue \fi

\newcommand{\fullonly}[1]{\iffullversion#1\else\expandafter\ignorespaces\fi}
\newcommand{\confonly}[1]{\iffullversion\expandafter\ignorespaces\else#1\fi}

\iffullversion
\includecomment{fullversion}\excludecomment{confversion}
\else
\excludecomment{fullversion}\includecomment{confversion}
\fi

\fullversiontrue

\title{\name: Breaking Location Privacy by Linking Device Identifiers Across Protocols}
\fullonly{\subtitle{\endgraf\vspace{2pt}\noindent\normalfont\normalsize\itshape
  This is the full version of the conference paper published in the proceedings of ACM CCS 2026.\\
  This version includes extended appendices. Please cite the conference version.\endgraf}}

\author{Aneet Kumar Dutta}
\affiliation{%
  \institution{CISPA Helmholtz Center for Information Security}
  \city{Saarbrucken}
  \country{Germany}}
\email{aneet.dutta@cispa.de}

\author{Mihirraj Dixit}
\affiliation{%
  \institution{Saarland University}
  \city{Saarbrucken}
  \country{Germany}}
\email{dixitmn6@gmail.com}

\author{Kevin Gni}
\affiliation{%
  \institution{CISPA Helmholtz Center for Information Security}
  \city{Saarbrucken}
  \country{Germany}}
\email{bin.ni@cispa.de}

\author{Wouter Lueks}
\affiliation{%
  \institution{CISPA Helmholtz Center for Information Security}
  \city{Saarbrucken}
  \country{Germany}}
\email{lueks@cispa.de}

\author{Mridula Singh}
\affiliation{%
  \institution{CISPA Helmholtz Center for Information Security}
  \city{Saarbrucken}
  \country{Germany}}
\email{singh@cispa.de}

\confonly{\settopmatter{printfolios=true}}

\confonly{\ccsdesc[500]{Security and privacy}}
\confonly{\ccsdesc[500]{Security and privacy~Mobile and wireless security}}

\confonly{\keywords{location privacy, mobile privacy, cross protocol}}

\setcopyright{acmlicensed} 
\copyrightyear{2026} 
\acmYear{2026} 
\acmDOI{XXXXXXX.XXXXXXX} 
\acmConference[CCS'26]{the 2026 ACM SIGSAC Conference on Computer and Communications Security}{November 15--19,2026}{The Hague, Netsherlands.}
\acmBooktitle{Proceedings of the 2026 ACM SIGSAC Conference on Computer and Communications Security (CCS'26), November 15--19,2026, The Hague, Netherlands.
}
\acmISBN{978-1-4503-XXXX-X/2018/06}  

\confonly{\settopmatter{printacmref=true}}
\fullonly{\settopmatter{printacmref=false}\renewcommand\footnotetextcopyrightpermission[1]{}}
\begin{document}

\begin{abstract}
Smartphones simultaneously transmit temporary identifiers over LTE, WiFi, and BLE. Existing privacy defenses analyze identifier randomization per protocol, implicitly assuming that these protections compose across protocols. We show that they do not: Even when each protocol leaks only temporary identifiers and the adversary is fully passive, unsynchronized identifier rotations allow cross-protocol stitching of device traces. We present \name, an uncertainty-aware tracing algorithm that links identifiers across time, space, and protocols under noisy localization and mobility. We evaluate \name using controlled lab experiments with commodity devices and large-scale mobility simulation. Under large-scale mobility simulation, \name reconstructs full traces for 83\% of users, versus 22\% for the best single-protocol baseline, showing that location privacy must be analyzed jointly across protocols. We further show that \name remains effective under partial coverage: strategically placed sniffers near LTE handover regions, mobile sniffers, and limited high-coverage subregions retain sufficient cross-protocol evidence to bridge observation gaps, achieving substantially higher linkability than random deployments.

\end{abstract}


\maketitle
\fullonly{\pagestyle{plain}}

\section{Introduction}

In wireless communication protocols, identifier randomization is the primary defense against passive device tracking. Early protocol implementations exposed static identifiers, thus eavesdroppers could trivially track the devices across time and space~\cite{lte_tracking_1_imsi, GruteserG05}. Modern protocol implementations address this through identifier rotation: WiFi and Bluetooth Low Energy randomize MAC addresses~\cite{ieee_wifi,ieee.802.11.2020}, while LTE replaced long-term subscriber identities with temporary values such as TMSI and C-RNTI~\cite{3gpp136300_ho}. 

Prior work shows that per-protocol randomization often fails in practice: attackers can defeat MAC randomization using fingerprints~\cite{ble_phy, wifi_mac}, by extracting static information to link the randomized identifiers~\cite{trackingble, asimov, ltrack}, or through specific implementation flaws that enable combining multiple aspects of a single protocol~\cite{ltrack,blc_linking,breaking_layer_two}. 
Even when identifiers are randomized, linkage can occur if other fields in the same transmission rotate on a different schedule~\cite{trackingble}. The Google/Apple Exposure Notification (GAEN) framework~\cite{GAEN} recognized this risk: it explicitly synchronized rotation of the BLE MAC addresses and the identifiers contained within broadcasts, so that they cannot be linked across advertisements. 

Modern smartphones are multi-protocol devices and thus present a harder problem: they emit identifiers simultaneously across LTE, WiFi, and BLE protocols developed by separate standards bodies, operating on independent schedules/triggers, with no mechanism to coordinate rotation. Prior privacy analyses have largely studied each protocol in isolation, implicitly assuming that if each individual protocol reveals only short-lived identifiers, tracking is confined to their brief lifetimes.

We ask whether this assumption holds in multi-protocol setting when all protocols \emph{work as intended}: no fingerprinting, no static identifiers, no protocol-specific flaws, and no active manipulation: Can a \emph{fully passive} eavesdropper still track a device using only \emph{temporary} identifiers observed across protocols?

The answer is yes. The principle is the same one that motivated GAEN: unsynchronized rotation creates linkage opportunities. In practice, identifier rotations across protocols are rarely synchronized: when one identifier changes, others often remain stable, letting an adversary bridge rotations via unchanged identifiers. Repeated over time, these bridges convert short unlinkable fragments into extended traces. This is not an implementation flaw to be patched; it is a structural consequence of per-protocol design.

However, exploiting this structure \emph{at scale} is non-trivial. Passive adversaries cannot trigger transmissions, force reconnects, or manipulate network behavior~\cite{ltrack,sigover1,adaptover,shaik_2016,lte_tracking_1_imsi}, so they must infer linkage from infrequent, asynchronous observations subject to localization errors and user mobility. Therefore, no single observation typically isolates a device; cross-protocol tracking is, instead, a spatiotemporal inference problem under uncertainty.

We present \name, an uncertainty-aware tracing attack that links temporary identifiers across time, space, and protocols using only passive observations. \name first constructs permissive intra-protocol and inter-protocol candidate links using temporal overlap, coarse localization, and mobility constraints, and  then iteratively prunes inconsistent candidates using cross-protocol evidence. It reconstructs traces only when the remaining chains are unambiguous. 

We validate \name through controlled experiments on commodity devices, micro-deployments, and large-scale simulations parameterized by real measurements. The micro-deployment confirms that end-to-end cross-protocol linkage is feasible in practice. Simulations for dense urban settings show that \name reconstructs complete observed traces for about 83\% of users, compared with about 22\% for the best single-protocol baseline. These results show that observing multiple protocols substantially increases the feasibility of passive tracking, even when each protocol individually deploys temporary identifiers.

These results reveal a structural privacy gap: identifier rotation protects privacy only when there exists ambiguity about which post-rotation identifier corresponds to which pre-rotation identifier. For multi-protocol devices, this ambiguity must hold jointly across all monitored protocols. We capture this requirement via a mixing condition: over a common time window, devices must remain spatially indistinguishable and the identifier changes on every monitored protocol must leave the adversary unable to unambiguously link the rotation. As the number of observed protocols increases, satisfying this joint mixing condition becomes less probable.

In summary, we make the following contributions:

\begin{itemize}
\item We show that a \emph{fully passive} adversary can perform large-scale physical tracking by linking observations across time, space, and protocols.

\item We design \name, an algorithm that reconstructs device traces from asynchronous, noisy, multi-protocol observations while accounting for localization error and mobility.

\item We develop a mathematical model for when identifier rotation prevents passive tracking. The model shows that privacy requires devices to ``mix'' in physical space: after changing identifiers, they must become indistinguishable from nearby devices that also change identifiers.

\item We evaluate \name through both lab experiments on real devices and large-scale simulations parameterized by BLE, WiFi, and LTE measurements.

\item We evaluate multiple sniffer placement strategies and show that \name can achieve high privacy leakage without dense citywide infrastructure. Even a few mobile sniffers can reconstruct long traces of the wider population.

\end{itemize}

\section{Background and Related Work}

This section reviews the identifiers and rotation mechanisms in LTE, WiFi, and BLE, and prior tracking attacks. 

\smallskip
\noindent
\textbf{Cellular LTE:}
The International Mobile Subscriber Identity (IMSI) uniquely identifies a subscriber's SIM card for enabling cellular communication. Because IMSI exposure enables direct long-term tracking, IMSI-catchers and signal-overshadowing attacks have received significant attention~\cite{lte_tracking_1_imsi,shaik_2016,privacy_public_safety,adaptover,ltrack}. To mitigate the tracking risk, 3GPP introduced the Temporary Mobile Subscriber Identity (TMSI), which the network can rotate during location updates and radio link failures~\cite{3gpp.23.003,3gpp.29.118}.
In practice, the most frequently observed identifier is the Cell Radio Network Temporary Identifier (C-RNTI), used for radio resource management and transmitted during connection establishment, maintenance, and release~\cite{3gpp.36.321}. A passive adversary can recover C-RNTI from over-the-air transmissions~\cite{ltrack,lte_sniffer_1}. The network rotates C-RNTI during inter-eNodeB handovers, radio link failures, and connection re-establishment~\cite{3gpp.36.331.18.5.0}.

Prior work has demonstrated several ways to link these temporary identifiers. Rupprecht et al.~\cite{breaking_layer_two} and Bae et al.~\cite{Bae22_watchingthewatcher} show that MAC and RRC layer information can associate C-RNTI and TMSI during RRC connection establishment -- though this is limited to the initial setup phase. Jover et al.~\cite{jover2016lte} observe C-RNTI changes during handover, but this linkage is unavailable once handover messages are protected on SRB-1 after security activation. Kotuliak et al.~\cite{ltrack} enable long-term tracking by binding the permanent IMSI to the temporary TMSI, but this requires active attacks to obtain the IMSI.

\smallskip
\noindent
\textbf{WiFi:} WiFi devices transmit MAC addresses in management, control, and data frames~\cite{ieee_wifi}. To prevent tracking, modern devices randomize MAC addresses, but implementations vary across operating systems due to the lack of a standardized specifications. This inconsistency has enabled fingerprinting attacks that defeat randomization. Vanhoef et al.~\cite{wifi_mac} fingerprint devices using probe-request contents and cryptographic weaknesses. Matte et al.~\cite{wifi_timing} exploit device-specific probe-request timing. Ribeiro et al.~\cite{asimov} combine fingerprints with RSSI-based localization for tracking.

\newcolumntype{L}[1]{>{\raggedright\arraybackslash}p{#1}}
\newcolumntype{C}[1]{>{\centering\arraybackslash}p{#1}}
\newcolumntype{Y}{>{\raggedright\arraybackslash}X}

\newcommand{\hdr}[2]{\parbox[c]{#1}{\centering\textbf{#2}}}

\begin{table}[t]
\centering
\footnotesize
\setlength{\tabcolsep}{2.5pt}
\renewcommand{\arraystretch}{1.08}
\caption{Overview of related tracking techniques across LTE, WiFi, and BLE.}
\label{tab:tracking_related}

\begin{tabularx}{\columnwidth}{@{}X@{}c c c c c@{}}
\hline
\textbf{Work} &
\textbf{Protocols} &
\shortstack{Fully\\pass.} &
\shortstack{Temp.\\IDs only} &
\shortstack{No finger./\\Static IDs} &
\shortstack{No prot.\\ flaw} \\
\hline

IMSI-catchers \cite{lte_tracking_1_imsi,adaptover,privacy_public_safety,ltrack} & LTE
& \xmark & \xmark & \xmark & \xmark \\

TMSI–C-RNTI map \cite{breaking_layer_two,Bae22_watchingthewatcher} & LTE
& \cmark & \cmark & \cmark & \xmark \\

C-RNTI tracking 
\cite{jover2016lte} & LTE
& \cmark & \cmark & \cmark & \xmark \\
\hline

Probe fingerprinting \cite{wifi_mac} & WiFi
& \cmark & \xmark & \xmark & \xmark \\

RSSI+fingerprint \cite{wifi_timing,asimov} & WiFi
& \cmark & \xmark & \xmark & \xmark \\

Side channel 
\cite{ndss_deanonymization_ble_wifi} & WiFi/BLE
& \xmark & \cmark & \cmark & \xmark \\
\hline

Device identifying token-BLE MAC linking
\cite{trackingble} & BLE
& \cmark & \xmark & \xmark & \xmark \\

BTC–BLE linking \cite{blc_linking} & BTC+BLE
& \cmark & \xmark & \xmark & \xmark \\

Phys.-layer fingerprint
\cite{ble_phy} & BLE
& \cmark & \xmark & \xmark & \cmark \\
\hline

\textbf{\name (Ours)} & \textbf{LTE+WiFi+BLE}
& \cmark & \cmark & \cmark & \cmark \\
\hline

\end{tabularx}
\end{table}

\smallskip
\noindent
\textbf{Bluetooth:} 
Bluetooth Classic uses a static, globally unique Bluetooth Device Address (BDADDR), while Bluetooth Low Energy (BLE) periodically randomizes this identifier~\cite{bluetoothprivacy}. Despite randomization, BLE remains vulnerable to tracking. Becker et al.~\cite{trackingble} show that advertisement contents enable fingerprint-based tracking. Ludant et al.~\cite{blc_linking} link static Classic BDADDR to dynamic BLE addresses by exploiting temporal correlations when both are implemented on the same chip. Givehchian et al.~\cite{ble_phy} track devices using hardware-level PHY characteristics. Ellis et al.~\cite{ndss_deanonymization_ble_wifi} demonstrate that paired devices leak side-channel information through distinguishable success/failure traffic patterns, enabling deanonymization even with MAC randomization. Jouans et al.~\cite{inria_ble} demonstrate that BLE identifiers can be temporally correlated in co-located spaces, though without considering mobility or localization uncertainty.

\smallskip
\noindent\textbf{Localization:}
Tracking requires localizing the identifiers observed over the air, and each protocol offers different localization capabilities. In BLE and WiFi, a passive adversary uses physical-layer measurements such as RSSI to estimate proximity~\cite{ble_loc_1,ble_loc_2,ble_loc_3,ble_loc_4,ble_loc_5,wifi_loc_1,wifi_loc_3}. In LTE, Timing Advance (TA) commands reveal the distance between the device and the base station. Roth et al.~\cite{location_privacy_lte} use TA for coarse range estimation; Kotuliak et al.~\cite{ltrack} refine this by combining TA with timing observations. These methods create protocol-dependent uncertainty that \name must account for.

\smallskip
As~\autoref{tab:tracking_related} shows, prior tracking attacks operate within a single protocol, linking observations via static identifiers, fingerprints, implementation flaws, or active manipulation, implicitly assuming that temporary identifier rotation bounds passive tracking to the temporary identifier lifetime. Modern devices break this assumption at the system level: identifiers rotate independently across protocols, and even GAEN~\cite{GAEN} required explicit synchronization within one ecosystem. \name exploits this composition gap, linking temporary identifiers across protocols without static information, fingerprints, protocol-specific flaws, or active interaction.

\section{Threat Model \& Attacker Goal}

We consider a passive adversary that aims to reconstruct device location traces from wireless observations, thereby violating the privacy of the people carrying those devices. The adversary eavesdrops on device transmissions, including infrastructure communication (e.g., with base stations or access points) and protocol broadcasts. It does not inject, relay, jam, or modify traffic, and it does not compromise tracked devices or network infrastructure.

We assume the attacker controls a set of sniffers geographically distributed within the target area. Sniffers may monitor one or more wireless protocols (e.g., LTE, WiFi, BLE), and sniffers for different protocols need not be co-located. We consider (i) \emph{fixed} sniffers deployed at known locations and (ii) \emph{mobile} sniffers carried by colluding/compromised devices that only passively record locally received transmissions. The placement and coverage of sniffers depends upon the adversary's budget and strategy.

Each sniffer can estimate the distance to a transmitting device using protocol-specific measurements: RSSI for WiFi and BLE, timing advance for LTE. We model each observation as constraining the device to a disk centered on the sniffer, with radius determined by measurement error. \rev{We characterize these protocol-specific measurement errors from prior works in Section~\ref{section:characterization} (Table~\ref{tab:protocol-summary}) and model them under a distance-dependent heavy-tailed and bounded error regime in Section~\ref{sec:simulation_setup}, evaluating \name's robustness to them in Section~\ref{sec:evaluation}.} More sophisticated techniques (multilateration, angle of arrival) would tighten these constraints; our results are conservative in assuming only range estimates.

Following prior work~\cite{reza_loc}, we focus on recovering location traces for devices and do not aim to link these traces to individuals. It is well known that given location traces, it is relatively easy to link them to individuals~\cite{FreudigerSH11}: e.g., by using specific points of interest such as home or work addresses~\cite{Krumm07}. Our contribution is showing that such traces can be reconstructed using temporary identifiers.

\smallskip
\noindent
\textbf{Effectiveness of the Attack.}
To capture the effectiveness of the attacker, we measure the maximum duration during which an attacker can recover a location trace of a specific device. To obtain a privacy metric, we normalize by the maximum duration that the attacker can observe the device.

\begin{definition}
   The location privacy leakage $\textsf{LP}_{\mathcal{D}}$ of a device $\mathcal{D}$ that is being observed by an attacker $\mathcal{A}$ (observations can come from different sniffers) recovering partial traces $\{ \textsf{trace}_i \}_i$ is defined as:
   \begin{equation*}
     \textsf{LP}_{\mathcal{D}} = \frac{\textrm{Length of longest $\textsf{trace}_i$ matching device $\mathcal{D}$}}{\textrm{duration during which device $\mathcal{D}$ was observed by $\mathcal{A}$}}
   \end{equation*}
   \label{def:privacy}
 \end{definition}
This definition simplifies Shokri et al.'s probabilistic formulation~\cite{reza_loc} by ignoring prior knowledge and counting only correctly reconstructed traces. We do not require the attacker to attribute traces to specific devices, since prior techniques already address this~\cite{FreudigerSH11,Krumm07}; our goal is to assess the correctness of trace recovery in the first place. $\textsf{LP}_{\mathcal{D}}$ captures the privacy failure that randomization aims to prevent. Identifier rotation should create ambiguity at rotation events; privacy fails when the adversary removes this ambiguity and tracks continuously across rotations. Thus, $\textsf{LP}_{\mathcal{D}}$ measures the fraction of a device's observable trajectory that the adversary follows continuously and unambiguously.

\section{Motivation and Technical Challenges}

Existing protocol implementations~\cite{ieee_wifi,ieee.802.11.2020,3gpp136300_ho} assume temporary identifiers bound the adversary's tracking window: once an identifier rotates and mixes with others, the trace ends. We show that this assumption fails for multi-protocol devices. A fully passive adversary can track devices far beyond any single protocol's rotation interval by exploiting unsynchronized temporary identifiers across LTE, WiFi, and BLE, without using IMSI catchers, protocol manipulation, or active interaction.

\begin{figure}[tbp]
    \centering
    \includegraphics[width=0.75\linewidth]{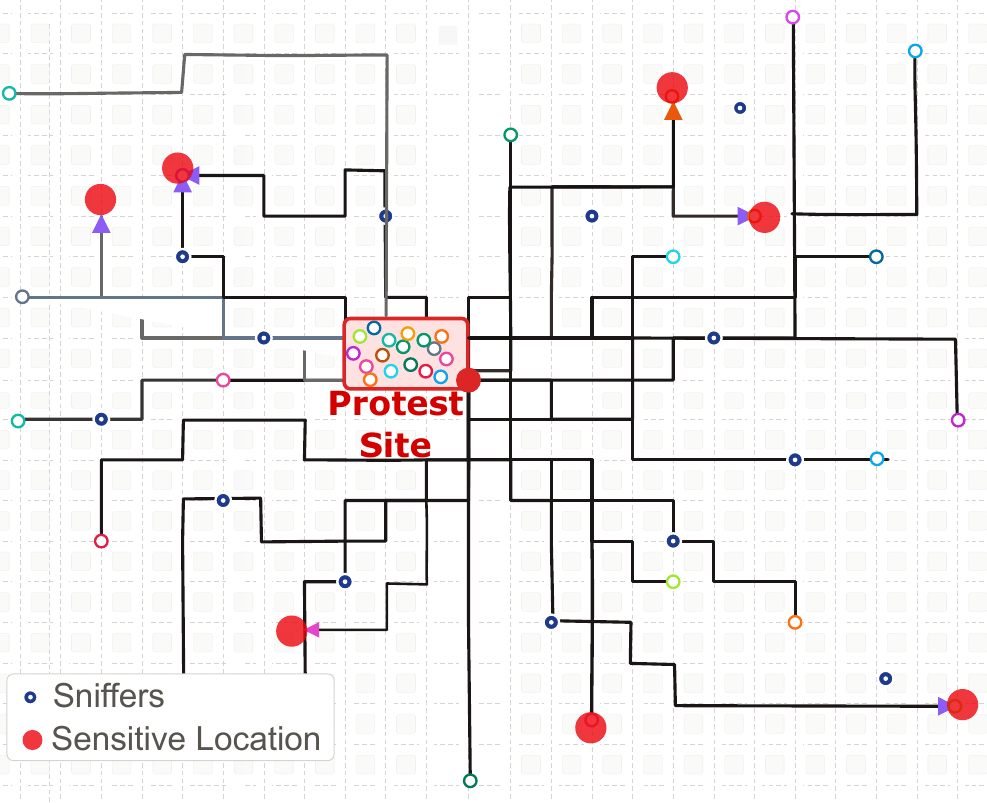}
    \caption{Example scenario: Tracking protest attendees to infer sensitive locations such as home or work addresses.}
    \Description{Example protest scenario}
    \label{fig:protest}
\end{figure}

\subsection{Tracking Protesters: An Example Scenario.}
\label{sec:motivation}

Consider a political protest in an urban area. A state adversary deploys passive sniffers near the protest site and at select city locations (Figure~\ref{fig:protest}). A device observed at the protest becomes a target, and temporary identifiers should make it unlinkable after rotation.

Cross-protocol observations defeat this protection. LTE, WiFi, and BLE rotate independently, so a stable identifier in one protocol can bridge rotations in another and let the adversary follow the device after it leaves the protest. The resulting harm goes beyond longer linkability: protest attendance can be tied to later movement, including home or work locations that may identify users~\cite{FreudigerSH11,Krumm07}, exposing participants to legal scrutiny, retaliation, or harassment.

\begin{figure}[tbp]
    \centering
    \includegraphics[width=0.75\linewidth]{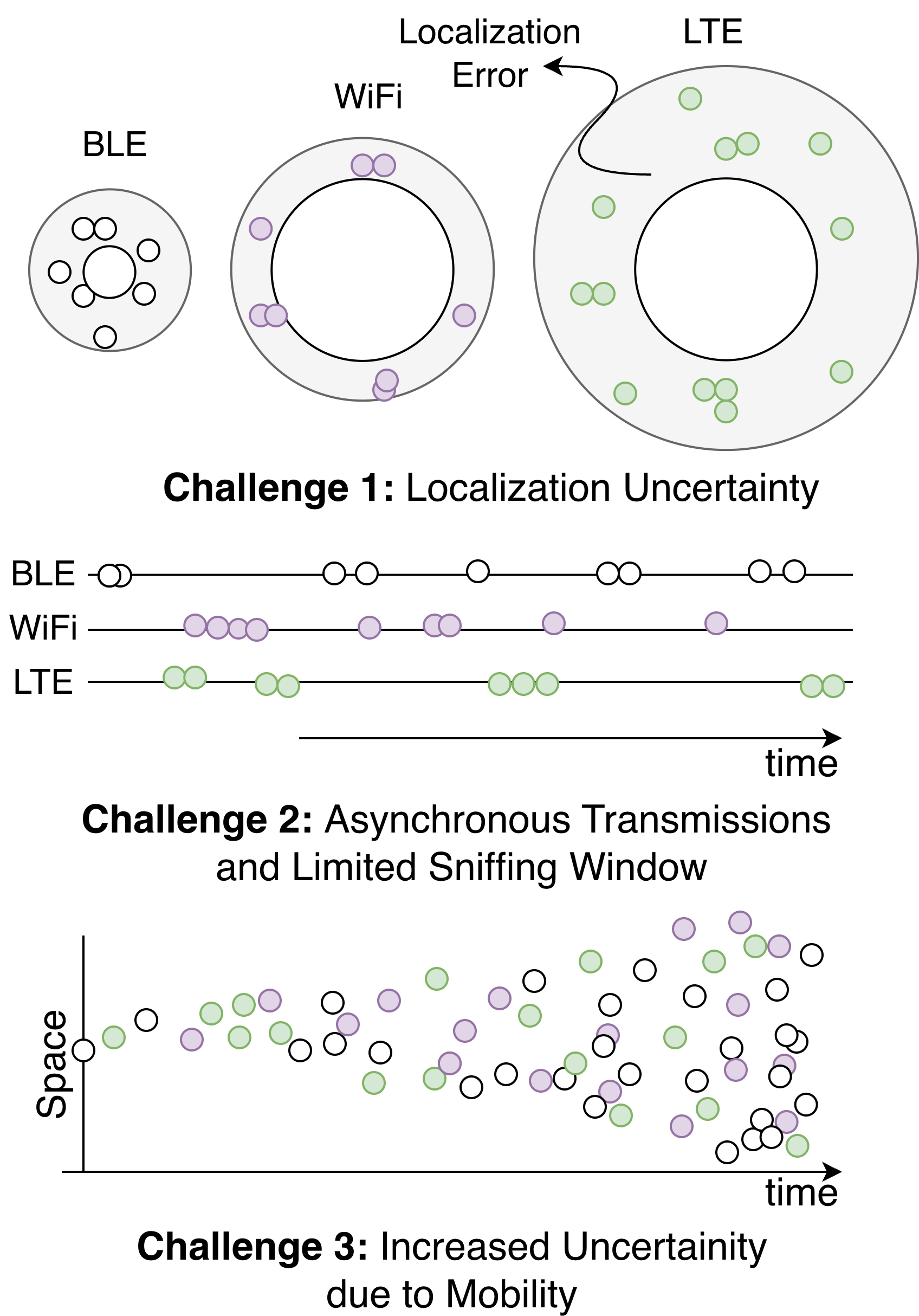}
    \caption{Challenges in linking identifiers across protocols.}
    \Description{Challenges}
    \label{fig:challenges}
\end{figure}

\subsection{Challenges in Cross-Protocol Linking}
\label{sec:challenges}
Although cross-protocol linking is conceptually simple, reliable large-scale tracking in dense environments is difficult because a fully passive adversary must overcome three challenges that simple co-occurrence checks cannot address:
\smallskip

\noindent\textbf{Challenge 1: Localization Uncertainty.} Sniffers provide imprecise location observations. They estimate distance using RSSI (WiFi, BLE) or timing advance (LTE)  with protocol-dependent errors. In crowded areas, these uncertainty regions overlap: one BLE identifier may match many nearby LTE devices, creating many plausible but false associations.

\smallskip
\noindent\textbf{Challenge 2: Asynchronous transmissions.} 
A passive adversary cannot trigger transmissions or control when devices emit. LTE, WiFi, and BLE transmit on independent schedules, so same-device identifiers are rarely observed simultaneously across protocols. The adversary therefore cannot rely on timing alone to establish cross-protocol links.

\smallskip
\noindent\textbf{Challenge 3: Mobility.}
Devices move between observations. A pedestrian walking at 1.5 m/s can travel 45 meters in thirty seconds, expanding the feasible region for the next transmission. In crowds, overlapping feasible regions create associations that are individually plausible but jointly inconsistent.

\begin{figure}[t]
    \centering
    \includegraphics[]{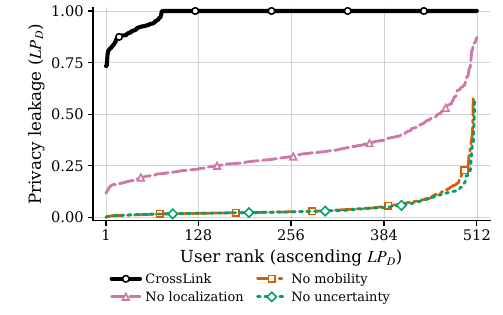}
    \caption{Effect of uncertainty modeling. \name recovers longer traces leading to higher privacy leakage. Ablated variants that ignore these uncertainties are less effective.}
    \label{fig:privacy_leakage_abalation}
    \Description{Ablation study}
\end{figure}

Uncertainty handling is therefore central to \name. Figure~\ref{fig:privacy_leakage_abalation} (detailed setup in Section~\ref{sec:sim_eval}) shows that ablated variants without uncertainty modeling recover far fewer traces than \name, confirming that uncertainty modeling is a prerequisite rather than a robustness improvement.

\section{Protocols Characterization}
\label{section:characterization}
In this section, we characterize the behavior of temporary identifiers across LTE, WiFi, and BLE under our passive-adversary threat model. Our measurements focus on three properties: the frequency of identifier-bearing transmissions, the availability of protocol-specific physical-layer signals for coarse localization, and the timing and conditions of identifier rotation.

We define an \emph{observation} as a tuple of a protocol-specific temporary identifier and a localization measurement: MAC + RSSI for WiFi and BLE; C-RNTI/TMSI + Timing Advance for LTE. We summarize our findings in Table~\ref{tab:protocol-summary}. Figures~\ref{fig:ti_lte_combined}--\ref{fig:ti_wifi_combined} report the distribution of transmission interval and randomization interval.

\subsection{Measurement Setup}
We measured transmission and identifier-rotation behavior on commodity smartphones: Samsung Galaxy M11, Xiaomi Redmi K50i, OnePlus Nord, Google Pixel 9, and iPhone 15. For each protocol, we passively captured over-the-air traffic under two operating modes: \emph{inactive} (screen off, no user interaction) and \emph{active} (calls, messaging, browsing). For LTE, we deployed an srsRAN-based eNodeB~\cite{srsran} on a USRP X310 with an Octoclock-G timing reference, and used a second USRP X310 as a passive sniffer~\cite{ltesniffer}. We cross-validated the main LTE observations with an Amarisoft Callbox~\cite{amarisoft_callbox}. For WiFi, we
captured traffic using an ALFA AWUS036ACM adapter with \texttt{hcxdumptool}. For BLE, we captured advertisements using an nRF52840 dongle. For each device and setting, we collected approximately two hours of data.

\begin{table}[t]
\centering
\caption{Protocol characteristics relevant to \name.
\rev{The Localization Error (LE) values denote the baseline uncertainty range per protocol informed by prior works. 
These are \emph{not} maximum-error bounds.} LP denotes the localization primitive. Transmission Interval (TI) and rotation behavior come from our measurements.}
\label{tab:protocol-summary}

\small
\setlength{\tabcolsep}{2pt}
\renewcommand{\arraystretch}{1.08}

\begin{tabularx}{\columnwidth}{
    @{}
    >{\raggedright\arraybackslash}p{0.24\columnwidth}
    >{\centering\arraybackslash}p{0.10\columnwidth}
    >{\centering\arraybackslash}p{0.13\columnwidth}
    >{\centering\arraybackslash}p{0.14\columnwidth}
    >{\raggedright\arraybackslash}X
    @{}
}
\toprule
\textbf{Protocol} & \textbf{LP} & \textbf{LE}
& \textbf{TI} & \textbf{Rotation behavior} \\
\midrule
LTE & TA & $ \pm 10$\,m & $0$--$60$\,s & Event-driven; mean $\sim 7$\,min \\

WiFi (connected) & RSSI & $ \pm 5$\,m & $0$--$20$\,s
& Stable over association \\

WiFi (disconnected) & RSSI & $ \pm 5$\,m & $1$--$5$\,min
& Vendor-dependent; often per probe \\

BLE & RSSI & $\pm 1.5$\,m & $0$--$40$\,s & Timer-based; $7$--$20$\,min \\

\bottomrule
\end{tabularx}
\end{table}

\subsection{Key Findings}
\label{sec:key-findings}
\noindent\textbf{Observation 1: Localization is protocol-dependent.}
In LTE, Timing Advance is always sent over plain-text revealing propagation delay that provides coarse range information to the serving base station. TA availability and update frequency depend on network configuration and implementation, with typical intervals ranging from 500\,ms to 10\,s~\cite{inactivity_timer,patent:inactivity_timer_nokia,3gpp32125}. In our srsRAN-based eNodeB testbed, the default configuration issues TA commands roughly every 0.5\,s (Appendix, Figure~\ref{fig:timing_advance}). \rev{We use distance error measurements from prior work. Kotuliak et al.~\cite{ltrack} report 90th-percentile distance errors of $4.672$--$7.238$\,m across four commercial phones. The work separately reports an error of $10.474$\,m for a USRP B210 running \texttt{srsUE} as an LTE UE.}

\rev{Localization accuracy using BLE and WiFi for a passive observer depends on the accuracy of RSSI measurements. Under favorable short-range line-of-sight conditions, passive RSSI observations can provide meter-level localization accuracy~\cite{ble_loc_1,ble_loc_2,ble_loc_3,ble_loc_5,ble_loc_4,wifi_loc_1,wifi_loc_3}. Nu{\ss}baumm{\"u}ller et al.~\cite{nussbaummuller_ble} evaluate smartphone BLE ranging over indoor and outdoor settings. Their baseline reports an overall RMSE of $1.20$\,m. Bao et al.~\cite{Bao2022} evaluate passive outdoor WiFi ranging from probe requests. Their  RSSI-to-distance models report MAEs of $3.44$--$4.62$\,m. In realistic urban environments, however, multipath propagation, interference, shadowing, and mobility produce both larger errors and violations of the expected monotonic relationship between RSSI and distance. Bao et al.~\cite{Bao2025Streetscape} show WiFi deployment on downtown streets can have  mean localization errors of up to 4.21\,m, with only 66--73\% of estimates within 4\,m and 17--21\% of observations discarded because of poor signal quality. Etzlinger et al.~\cite{etzlinger_ble_distance} evaluate commodity smartphones over $1$--$6$\,m across different carry positions, orientations, and propagation environments. They report an overall BLE ranging RMSE of $3$\,m.  These works establish that localization error due to inaccurate RSSI measurements can vary significantly in different settings.}

\smallskip
\noindent\textbf{Observation 2: Devices transmit frequently.}
Devices frequently transmit temporary identifiers across LTE, WiFi, and BLE. In LTE, UEs emit temporary identifier-bearing traffic even when inactive. Scheduled control/data messages exposing C-RNTI and TMSI appear during RRC connection establishment. After setup, UEs remain in \texttt{RRC\_CONNECTED} until the inactivity timer expires. Background services, notifications, calls, or data keep connections alive or trigger Service Requests. RRC measurement reports and MAC Buffer Status/Power Headroom Reports~\cite{3gpp32125,3gpp136304} expose C-RNTI traffic frequently (<30\,s) and more often when active.

In WiFi, transmission behavior depends strongly on the association state (connected/disconnected). When a device is inactive but remains associated with an access point, it continues to transmit control, ACK, and QoS packets roughly every 1--20\,s. Disconnected devices send probe requests in bursts at vendor-dependent intervals ranging from under 1 minute to 5 minutes. In BLE, devices transmit frequently, with intervals below 35\,s in both inactive and active mode. Prior work reports higher BLE transmission rates~\cite{inria_ble}, especially when specific applications (e.g., contact tracing applications) are installed~\cite{ble_phy}.

\smallskip
\noindent\textbf{Observation 3: Identifier rotations are protocol-dependent and unsynchronized.}
Identifier rotation happens under different triggers or conditions in LTE, WiFi, and BLE, and therefore rarely align across protocols. In LTE, passive C-RNTI/TMSI linking depends on the signaling bearer. Unprotected SRB-0 exposes C-RNTI--TMSI links during \texttt{RRCConnectionEstablishment}~\cite{breaking_layer_two,Bae22_watchingthewatcher} and old/new C-RNTI links during \texttt{RRCConnectionReestablishment} (Figs.~\ref{fig:rrc_reestablishment},~\ref{fig:test_rrc_connection_request}). Inter-eNodeB handovers assign a new C-RNTI through SRB-1-protected \texttt{RRCConnection\allowbreak Reconfiguration} with \textit{mobilityControlInfo}, so passive linking usually fails. Intra-eNodeB handovers may also use \textit{mobilityControlInfo}, but preserve the C-RNTI to maintain RRC context and UE continuity on handover failure~\cite[10.1.2.1]{3gpp136300_ho}; same-cell reconfigurations without \textit{mobilityControlInfo} also preserved it (Figs.~\ref{fig:rrc_connection_reconfig_no_change},~\ref{fig:rrc_intra_enb}). Thus, LTE-only passive tracking is unreliable across inter-eNodeB handovers. From repeated GPS-tagged pedestrian LTE logs recording serving PCI/base-station identity, we estimate inter-eNodeB handovers every $\sim$7\,min on average.

In WiFi, rotation behavior is state-dependent. MAC addresses remain stable throughout an association. In the disconnected state, the randomization behavior varies by vendor: some devices randomize at nearly every probe request, whereas others retain the same probe identifier for the entire period.

In BLE, MAC-address rotation happens in regular intervals. Across the measured Android and iOS devices, observed BLE randomization intervals fall roughly in the 7--20 minute range, consistent with the BLE specification~\cite{bluetoothprivacy}.

\smallskip
\noindent\textbf{Takeaway.}
Across LTE, WiFi, and BLE, devices provide noisy localization, asynchronous transmissions, and independent rotations, motivating \name design in the next section and parameterizing simulation in Section~\ref{sec:simulation_setup}.

\begin{figure*}[tbp]
    \centering
    \includegraphics[width=0.8\linewidth]{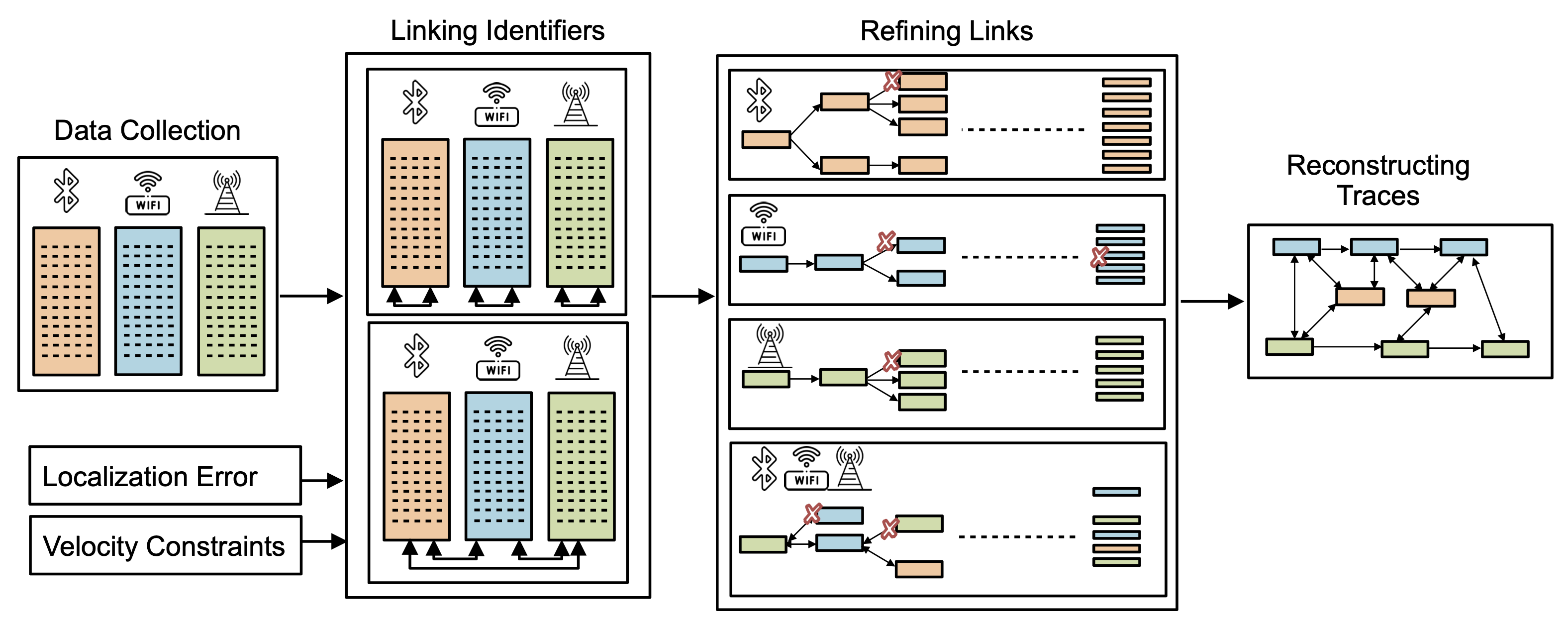}
     \caption{
    Overview of \name. \name builds permissive intra- and inter-protocol candidate links from noisy observations and mobility constraint, iteratively prunes them using cross-protocol consistency, and reconstructs traces from unique chains.}
    \label{fig:overview}
    \Description{Approach}
\end{figure*}

\section{Recovering Location Traces}
\label{sec:algorithm}

\noindent\textbf{Overview.} 
\name reconstructs device traces from asynchronous, noisy, multi-protocol observations in three phases (Figure~\ref{fig:overview}, Algorithm~\ref{alg:ltr}). (1) Candidate construction. \name builds permissive candidate links of two types using temporal overlap, spatial consistency, and mobility constraints: \emph{intra-protocol} links connect an identifier to its post-rotation successor within the same protocol; \emph{inter-protocol} links connect identifiers of different protocols active simultaneously on the same device. (2) Cross-refinement. \name iteratively prunes intra-protocol links by using inter-protocol evidence and vice versa, until convergence. (3) Trace reconstruction. \name follows only unambiguous links and unique chains to reconstruct traces.

\begin{algorithm}[tbp]
\caption{Finding Potential Identifier Maps}
\label{alg:ltr}
\footnotesize

\KwIn{$\data$ collected by all the sniffers for the entire duration, set of protocols $P$}
\KwOut{$\InterMap{.}{.}$, $\IntraMap{.}$}

$\InterMap{\cdot}{\cdot}:=\textit{ComputeInterMap}(\data,P)$;\

$\IntraMap{\cdot}:=\textit{ComputeIntraMap}(\data,P)$;\

Initialize $\InterMap{\cdot}{\cdot}^{\code{prev}}, \IntraMap{\cdot}^{\code{prev}}$ to empty maps\;

\While{$(\IntraMap{\cdot} \neq \IntraMap{\cdot}^{\code{prev}}) \lor \allowbreak (\InterMap{\cdot}{\cdot} \neq \InterMap{\cdot}{\cdot}^{\code{prev}})$}{
$\InterMap{\cdot}{\cdot}^{\code{prev}}, \IntraMap{\cdot}^{\code{prev}} := \InterMap{\cdot}{\cdot}, \IntraMap{\cdot}$;\

$\IntraMap{\cdot}=\textit{RefineIntraMap}(\IntraMap{\cdot},\InterMap{\cdot}{\cdot})$\;
$\InterMap{\cdot}{\cdot} =\textit{RefineInterMap}(\InterMap{\cdot}{\cdot},\IntraMap{\cdot})$\;

}

\Return $\IntraMap{.}$, $\InterMap{.}{.}$\
\end{algorithm}

\subsection{Linking Device Identifiers}

\newcommand{\vmax}{\ensuremath{V_{max}}}
\newcommand{\measerr}[1]{\ensuremath{\epsilon_{#1}}}

\newcommand{\sniffer}[1]{\mathcal{S}_{#1}}
\newcommand{\snifloc}[1]{\mathcal{S}_{#1}^{\mathrm{loc}}}

\newcommand{\maxiat}[1]{\code{max}_{\code{iat}}^{#1}}

We first determine which identifiers could potentially be linked. Our algorithms consider multiple types of protocols. Thus, we distinguish two types of links: \emph{Intra-protocol} links $\protidname{p_i}{1} \linksto \protidname{p_i}{2}$ connect successive identifiers of the same protocol $p_i$ on one device. \emph{Inter-protocol} links $\protid{p_i} \linksto \protid{p_{\!j}}$ connect identifiers of different protocols $p_i$ and $p_{\!j}$ active simultaneously on one device.

When identifying potential links, our algorithms are designed to \rev{construct permissive candidate sets to minimize false negatives}-- assuming we received corresponding observations. \rev{Therefore}, our initial linking process will likely result in many false positives that we filter out in the next cross-linking phase.

We assume a set of sniffers $\mathcal{S} = \{\mathcal{S}_1, \cdots, \mathcal{S}_{N}\}$. 
For each transmission that $\mathcal{S}_i$ receives, it records an observation $(\protid{}, p, t, d, \mathcal{S}_i)$, where $\protid{}$ is the protocol-$p$ identifier received at time $t$ and distance $d$ from the sniffer. Let $\mathcal{S}_{i}^{loc}$ be the location of $\mathcal{S}_{i}$.

\smallskip
\noindent\textbf{Transmissions from the same device.}
To identify potential intra- and inter-protocol links we first find observations $(\protid{}, p, t, d, \mathcal{S})$ and $(\protid{}', p', t', d', \mathcal{S}')$ of different identifiers $\protid{}$ and $\protid{}'$ that could have originated from the same device (for intra-links the protocols are the same, $p = p'$, while they are different for inter-links). 
Simply looking at the estimated location for these transmissions does not work. Devices move between transmissions, so $t \neq t'$ forces us to model mobility. Sniffers estimate distances, with protocol-specific error, so $d$ and $d'$ \rev{are noisy approximations of the actual distances}.

We make two assumptions about the observation model. First, the adversary knows an upper bound $\vmax$ on device speed. We use walking speed in most experiments and evaluate higher speeds in Section~\ref{para:q4_mobility}.

For an observation $(\protid{}, p, t, d, \mathcal{S})$, \rev{\name assumes that} the device lies anywhere at distance between $[d-\measerr{p},\, d+\measerr{p}]$ from sniffer $\mathcal{S}$. Here, $\measerr{p}$ is the protocol-specific \rev{error tolerance in distance estimation. If a range estimate differs from the true distance by more than its error tolerance, \name discards the true link.
Attackers would therefore choose a reasonable upper-bound on error tolerance ($\measerr{p}$). While a (too) large error tolerance causes more false positives to be included, \name filters many of these out again in the refinement phase.
We show in Section~\ref{sec:evaluation} that \name works as long as \measerr{p} is big enough to account for the actual measurement errors.}

Formally, observations $O = (\protid{}, p, t, d, \sniffer{i})$ and $O' = (\protid{}',\allowbreak p',\allowbreak t', d', \sniffer{j})$, are potentially transmitted by the same device if
\begin{equation}
\label{eqn:mob}
    |\snifloc{i} - \snifloc{j}| \leq (d+\measerr{p})+(d'+\measerr{p'})+|t-t'|\cdot V_{max}
\end{equation}
\name can incorporate stronger localization signals, such as angle of arrival, by replacing the range constraint with a tighter feasible region.

We also assume an upper-bound $\maxiat{p}$ on the inter-arrival time of messages for each protocol $p$. Our experiments show that such a maximum exists in practice. If we underestimate this bound, our algorithm will exclude potential links, and tracking of this identifier (for this protocol) will then fail.

\begin{algorithm}[tbp]
\footnotesize
\caption{Compute Inter-Protocol Mappings}
\label{alg:intermap_new}
\SetKwProg{Fn}{Function}{}{}

\Fn{ComputeInterMap($\data, P$)}{
\KwIn{Sample $\data$ collected by all the sniffers for the entire duration, set of protocols $P$}
\KwOut{A set of potential mappings $\InterMap{p_i}{p_j} : \{ \identifiers \textrm{ for protocol } p_i \} \to \mathcal{P}(\{ \identifiers \textrm{ for protocol } p_j) \})$}
Initialize empty map $\InterMap{\cdot}{\cdot}$\;

\ForEach{$(\id, p, t, d, S) \in \data$}
{
\ForEach{$p' \in P \setminus \{ p \}$}{
 $\InterMap{p}{p'}[\id] := \InterMap{p}{p'}[\id] \cup \allowbreak
 \{\id' \mid \allowbreak (\id', p', t', d', S') \in \data\allowbreak \land \textit{PotentialMatch}(\id, \id', p, p', \data)\}$
}
}

\Return $\InterMap{\cdot}{\cdot}$
}

\Fn{PotentialMatch($\id, \id', p, p', \data$)}{
\newcommand{\emptyparam}{\text{\textunderscore}}
\renewcommand{\emptyparam}{\cdot}

$\setidin{ID} = \{ O \mid O=(\id, p, \emptyparam, \emptyparam, \emptyparam) \in \data \}$ \; 
$\setidin{ID}^\prime= \{ O' \mid O'=(\id', p', \emptyparam, \emptyparam, \emptyparam) \in \data \}$ \; 
consistent := $\forall O \in \setidin{ID}, O' \in \setidin{ID}^\prime$: $O,O'$ satisfy eq.~\eqref{eqn:mob} \;
Let $t_f, t_l$ and $t_f', t_l'$ be the times of first and last observation in $\setidin{ID}$ respectively $\setidin{ID}^{\prime}$\;
overlap := $\neg \bigl[(t_l + \maxiat{p} < t_f') \lor (t_l' + \maxiat{p'} < t_f) \bigr]$\;
\Return consistent $\land$ overlap \;
}

\footnotesize
\end{algorithm}

\smallskip
\noindent\textbf{Identifying potential inter-protocol links.} Our algorithm first constructs candidate inter-protocol links. For each protocol pair $p \neq p'$, it seeks mappings $\protid{p} \linksto \protid{p'}$ between
identifiers that could belong to the same device. A mapping is feasible only if (1) all observations of $\protid{p}$ and $\protid{p'}$ satisfy the mobility and
localization constraints, and (2) the two identifiers were active during an overlapping time interval.

More formally, Algorithm~\ref{alg:intermap_new} builds the inter-protocol candidate map $\InterMap{p_i}{p_{\!j}}$. For each ordered pair of distinct protocols $(p_i,p_{\!j})$, it maps an identifier $\id$ from $p_i$ to identifiers from
$p_{\!j}$ that could belong to the same device. The algorithm iterates over identifiers $\id$ in protocol $p$ and other protocols $p'$ (lines~3--4), and adds $\id'$ to the candidate set when \textsc{PotentialMatch} succeeds (line~5).

\textsc{PotentialMatch} applies two conservative tests. First, it collects all observations of $\id$ and $\id'$ (lines~8--9) and requires every pair to satisfy the mobility constraint in Equation~\eqref{eqn:mob} (line~10). This rejects matches that cannot be co-located under the speed bound and localization \rev{error tolerance}; observations from multiple sniffers also impose multilateration constraints. Second, it computes the observed lifetimes of both identifiers (line~11) and requires them to overlap after accounting for protocol-specific inter-arrival
bounds (line~12). Pairs that pass both tests remain as inter-protocol candidates.

Because the feasibility test admits all links consistent with mobility and measurement-error bounds, $\InterMap{p_i}{p_j}$ contain false positives, which we prune in the refinement phase.

\smallskip
\noindent\textbf{Identifying potential intra-protocol links.}
Next, \name builds intra-protocol candidate links. For each protocol $p$, it seeks pairs $\id \linksto \id'$ where $\id'$ could be the randomized successor of $\id$. A candidate must satisfy two conditions: all observations of $\id$ and $\id'$ must be feasible for one moving device, and $\id'$ must appear soon after $\id$ disappears, within the protocol inter-arrival bound.

\begin{algorithm}[tbp]
\caption{ComputeIntraMap\label{alg:Intramap_new}}
\footnotesize

\SetKwProg{Fn}{Function}{}{}
\Fn{ComputeIntraMap($\data,P$)}{
    \KwIn{Sample $\data$ collected by all the sniffers for the entire duration}
    \KwOut{$\IntraMap{\cdot}$: \{$\identifiers$ of $p$\} $\linksto$ Potential(\{$\identifiers$ of $p$\})}
   Initialize empty map $\IntraMap{\cdot}$\;
    \ForEach{$p \in P$}{
    \ForEach{($\id,  p, \did, \tid, S)     \in \data$\label{alg:upintra:for-ids}}{
    $\IntraMap{p}[\id] := \IntraMap{p}[\id] \cup \allowbreak
 \{\id' \mid \allowbreak (\id', p, \cdot, \cdot, \cdot) \in \data\allowbreak \land \textit{PotentialMatch}(\id, \id', p, \data)\}$ \;
    }
    }
    \Return $\IntraMap{\cdot}$ 
}

\Fn{PotentialMatch($\id, \id', p, \data$)}{

\newcommand{\emptyparam}{\cdot}

$\setidin{ID} = \{ O \mid O=(\id, p, \emptyparam, \emptyparam, \emptyparam) \in \data \}$ \; 
$\setidin{ID}^\prime= \{ O' \mid O'=(\id', p, \emptyparam, \emptyparam, \emptyparam) \in \data \}$ \; 
consistent := $\forall O \in \setidin{ID}, O' \in \setidin{ID}^\prime$: $O,O'$ satisfy eq.~\eqref{eqn:mob} \;
Let $t_f, t_l$ and $t_f', t_l'$ be the times of first and last observation in $\setidin{ID}$ respectively $\setidin{ID}^{\prime}$\;
justafter := $t_l < t_s' \land (t_s' - t_l) < \maxiat{p}$\;
\Return consistent $\land$ justafter\;
}

\end{algorithm}

Algorithm~\ref{alg:Intramap_new} constructs $\IntraMap{p}$, which maps each identifier $\id$ to possible same-protocol successors. It iterates over protocols and identifiers (lines~3--4), and adds $\id'$ when \textsc{PotentialMatch} succeeds (line~5). \textsc{PotentialMatch} gathers all
observations of $\id$ and $\id'$ (lines~8--9), checks the localization and mobility constraint in Equation~\eqref{eqn:mob} (line~10), and verifies that $\id'$ starts soon after $\id$ ends (lines~11--12). This temporal test preserves the directed nature of identifier rotation: $\id$ may map to $\id'$, but not vice versa.

We are liberal in assigning potential intra-links due to the maximum mobility factor and the localization  \rev{error tolerance} which we refine in the next section. 
\subsection{Refining Links}

The potential inter and intra mappings that we constructed so far contain many false positives. Our key insight is that we can use information from inter-protocol links to refine intra-protocol links, and vice versa. Ideally, this process converges to a point where each identifier has a unique intra-protocol mapping, and each inter-protocol mapping of an identifier accurately reflects the identifiers from other protocols that were active during the identifier's lifetime on the same device.

To refine our potential mappings, we iteratively refine the intra- and inter-protocol links until no more refinement is possible (see lines 4--7 in Algorithm~\ref{alg:ltr}).

\smallskip
\noindent\textbf{Refining intra-protocol links.}
\name prunes same-protocol successor candidates using cross-protocol consistency. Suppose $\id'$ is the randomized successor of $\id$ in protocol $p$. Since protocols rotate independently, at least one identifier from another protocol often remains stable across this rotation. Thus, $\id$ and $\id'$ should share compatible inter-protocol evidence. \name discards a candidate link $\id \linksto \id'$ when their inter-protocol candidate sets have no common elements. When only one successor remains for $\id$, \name has found a true link, and removes that successor from competing predecessor sets.

\begin{algorithm}[tbp]
\footnotesize
\caption{RefineIntraMap}
\label{alg:refine-intra-map}

\SetKwProg{Fn}{Function}{}{}
\Fn{RefineIntraMap$(\IntraMap{\cdot}, \InterMap{\cdot}{\cdot}, P)$}{
\KwIn{Current $\IntraMap{\cdot}$ and $\InterMap{\cdot}{\cdot}$, protocols $P$}
\KwOut{Updated $\IntraMap{\cdot}$}
\ForEach{$p \in P$\label{alg:refintramap:for-prot}}{
\ForEach{$\id \in \IntraMap{p}$\label{alg:refintramap:for-id}}{
$S = \IntraMap{p}[\id]$\;\label{alg:refintramap:original}
$S_{\code{new}} = \{\id' \mid \id' \in S \land \allowbreak
  \forall p \neq p' \in P : \InterMap{p}{p'}[\id] \cap \InterMap{p}{p'}[\id'] \neq \varnothing $\}\;
  \label{alg:refintramap:refined}
$\IntraMap{p}(\id) := S_{\code{new}}$
}

}

$\IntraMap{}:=\textit{FilterIntramap}(\IntraMap{})$

\Return $\IntraMap{}$
}

\Fn{FilterIntramap($\IntraMap{}$)}{

\ForEach{$p \in P$}{
\ForEach{$\id \in \IntraMap{p}$}{
    \If{$| \IntraMap{p}[\id] | = 1$}{
      $\setidin{\id}=\IntraMap{p}[\id]$ \;
        \ForEach{$\id^{\prime} \in \IntraMap{}, \id^{\prime} \neq \id$}{
            $\IntraMap{p}[\id^{\prime}]:=\IntraMap{}[\id^{\prime}] \setminus \setidin{\id}$
        }
    }
}
}
\Return $\IntraMap{\cdot} $\;
}

\end{algorithm}

Algorithm~\ref{alg:refine-intra-map} formally applies this rule. It iterates over each protocol and identifier (lines~2--3), considers each candidate successor $\id' \in
\IntraMap{p}[\id]$ (line~4), and keeps $\id'$ only when $\id$ and $\id'$ share compatible inter-protocol set (line~5--6). After refining the identifiers it then invokes \textsc{FilterIntraMap} to propagate resolved rotations. \textsc{FilterIntraMap} scans all identifiers (lines~10--11), detects singleton successor sets (line~12), and removes each resolved successor from all other predecessor sets s(lines~13--15). This propagation turns unambiguous rotations into constraints for the remaining ambiguous ones.

\begin{algorithm}[tbp]
  \footnotesize
  \SetKwProg{Fn}{Function}{}{}
  \caption{RefineInterMap\label{alg:refine-inter-map}}
  \Fn{RefineInterMap($\InterMap{\cdot}{\cdot},\allowbreak \IntraMap{\cdot}, P$)}{
  \KwIn{Current $\IntraMap{\cdot}$ and $\InterMap{\cdot}{\cdot}$, protocols $P$}
  \KwOut{Updated $\InterMap{\cdot}{\cdot}$}

  \ForEach{$p$,$p'\in P$, $p\neq p'$\label{alg:refintermap:for-prots}}{
      \ForEach{$\id \in \InterMap{p}{p'}$\label{alg:refintermap:for-ids}}{
          $T = \IntraMap{p}[\id]$\;
         
          $\code{retain} := \InterMap{p}{p'}[\id] \;\cap\allowbreak \cup_{\id' \in T} (\InterMap{p}{p'}[\id'])$\;
          $\code{retain}_{\code{prev}} := \emptyset$\;
          \While{$\code{retain}_{\code{prev}} \neq \code{retain}$}{
              $\code{retain}_{\code{prev}} := \code{retain}$\;
              $\code{retain} := \code{retain}_{\code{prev}} \cup \{ \id' \;|\; \allowbreak
              \id' \in (\InterMap{p}{p'}[\id] \setminus \code{retain}_{\code{prev}}) \;\land\; 
                    (\IntraMap{p'}[\id'] \cap \code{retain}_{\code{prev}}) \neq \emptyset \} $\;
          }
         
          $\InterMap{p}{p'}[\id]:=\code{retain}$\;
          }

      }

      $\InterMap{\cdot}{\cdot}:=\textit{MakeSymmetric}(\InterMap{\cdot}{\cdot}, P)$

\Return $\InterMap{\cdot}{\cdot}$\; 
}

\Fn{MakeSymmetric($\InterMap{\cdot}{\cdot}, P)$}{

\ForEach{$p$,$p'\in P$, $p\neq p'$}{
\ForEach{$\id \in \InterMap{p}{p'}$}{
$\InterMap{\pid}{p}[\id]:=\{\id^{\prime} \;|\; \id^{\prime} \in \InterMap{p}{p'}[\id] \land \id \in \InterMap{p'}{p}[\id^{\prime}]\}$
}
}
\Return $\InterMap{\cdot}{\cdot}$
}

\end{algorithm}

\smallskip
\noindent\textbf{Refining inter-protocol links.}
The initial inter-protocol maps are intentionally permissive and therefore contain many false positives. \name prunes them using consistency across same-protocol rotations. Let
$c \in \InterMap{p}{p'}[\id]$ be a candidate identifier from protocol $p'$. If $c$ and $\id$ belong to the same device, then this relationship should remain consistent when $\id$ rotates to some successor $\id' \in \IntraMap{p}[\id]$. There are two cases. If protocol $p'$ does not rotate during this interval, then $c$ should also appear in $\InterMap{p}{p'}[\id']$. If protocol $p'$ does rotate, then one of $c$'s same-protocol successors should appear there instead. Otherwise, $c$ has no support across the rotation of $\id$, so \name removes
it.

Formally, Algorithm~\ref{alg:refine-inter-map} applies this rule. For each protocol pair and identifier (lines~2--3), it refines $\InterMap{p}{p'}[\id]$. It first keeps
the simple case: candidates that also appear in the inter-map of at least one successor $\id' \in \IntraMap{p}[\id]$ (line~5). These candidates remain stable while $\id$ rotates. This retention is robust to near-simultaneous rotations
because Algorithm~\ref{alg:intermap_new} already accounts for the maximum inter-arrival time when testing temporal overlap.

The algorithm then handles the case where protocol $p'$ rotates. Starting from the retained candidates, it repeatedly adds any predecessor that is both in $\InterMap{p}{p'}[\id]$ and linked by the intra-protocol map of $p'$ (lines~7--9).
This backward expansion preserves valid chains when $p'$ rotates more often than $p$. Finally, \textsc{MakeSymmetric} enforces bidirectional consistency (line~11). It scans protocol pairs and identifiers (lines~14--15) and keeps $c \in \InterMap{p}{p'}[\id]$ only if $\id \in \InterMap{p'}{p}[c]$ (line~16).
\subsection{Reconstructing Traces}

\begin{algorithm}[tbp]
\caption{Reconstructing Traces}
\label{alg:reconstruction}
\footnotesize
\SetInd{0.1em}{1em}
\SetKwProg{Fn}{Function}{}{}
\Fn{Reconstruct$(\id^*, p^*, \IntraMap{.}, \InterMap{.}{.}, \data)$}{
  \KwIn{Seed identifier $\id$ for protocol $p$, $\InterMap{.}{.}$, $\IntraMap{.}$ as per by Alg.~\ref{alg:ltr}, all observations $\data$}
  \KwOut{Longest sequence of identifiers produces by the same device that transmitted $\id$}

  $V := \{(\id, p) \mid (\id, p, \cdot, \cdot, \cdot) \in \data \}$\;
  $E_{\code{intra}} = \{ ((\id, p), (\id', p)) \mid (\id, p) \in V \land \allowbreak \IntraMap{p}[\id] = \{ \id' \} \}$\;
  $E_{\code{inter}} = \{ ((\id, p), (\id', p')) \mid (\id, p) \in V \land \allowbreak \id' \in \InterMap{p}{p'}[\id] \land
  \code{LieOnChain}(\InterMap{p}{p'}[\id], p', \IntraMap{\cdot}) \}$\;
  Let $G = (V, E_{\code{intra}} \cup E_{\code{inter}})$ be a directed graph where path lengths corresponds to total duration of the identifiers as per $\data$\;
  \Return longest path in $G$ originating at $(\id^*, p^*)$
}
\Fn{LieOnChain$(S, p, \IntraMap{\cdot})$}{
Let $S = \{\id_1, \ldots, \id_n \}$\;
\Return $\exists \pi: [n] \to [n] : \forall i \in [n - 1] : \IntraMap{p}[\id_{\pi(i)}] = \{ \id_{\pi(i + 1)} \}$
}

\end{algorithm}
At the end of Algorithm~\ref{alg:ltr}, we have computed and refined all potential inter and intra-protocol links. We can now use these to trace identifiers, as shown in Algorithm~\ref{alg:reconstruction}. We construct a graph, where the vertices $V$ consist of observed identifiers, and edges ($E_{\code{intra}}$ and $E_{\code{inter}}$) between these consist of all the links where there is exactly one consistent mapping. For $E_{\code{intra}}$ we enforce the consistency requirement by requiring that $\IntraMap{p}[\id]$ contains a singleton set. This does not work for inter mappings, however, as a single protocol $p$ identifier $\id$ might legitimately overlap with multiple protocol $p'$ identifiers. We therefore require instead that all identifiers $\InterMap{p}{p'}[\id]$ lie on a unique intra-mapping chain (see lines 7--9).

We define the length of paths on this graph as the difference in time between when the final identifier in the path was last observed and when the first identifier was first observed. The longest we can track a device that transmitted an identifier $\id^*$ is then given by the longest path on the graph.

\section{When Does Randomization Prevent Tracking?}
\label{sec:mixing}
To characterize the strongest protection identifier randomization can provide against a passive multi-protocol adversary, we bound the adversary's probability of failing to link rotations. We adapt mix zones from location privacy~\cite{BeresfordS03,BeresfordS04}: two devices mix for protocol $P$ and time window $T$ if the adversary cannot deterministically match post-rotation identifiers to devices. The window is a mix zone when the devices remain location-indistinguishable throughout $T$.

We analyze two devices and one protocol $P$. We assume (i) the devices remain location-indistinguishable for the entire duration $T$, (ii) each device’s transmission and randomization processes are independent of the other device, and (iii) protocol processes are independent across protocols.

Perfect mixing for protocol $P$ within duration $T$ requires two conditions:
\begin{enumerate}
   \label{cond:perfect_mixing}
    \item \textbf{Both devices rotate.} Each device must randomize its identifier at least once within $T$.
    \item \textbf{Transmission order.} After the first transmission of any \emph{new} identifier, the adversary must not observe an \emph{old} identifier from the other device before that device starts transmitting its own new identifier. This condition prevents deterministic linkage when one new identifier appears while the other device's old identifier remains visible.
\end{enumerate}

\newcommand{\expdistr}[1]{\mathrm{Exp}(#1)}
\newcommand{\paramrand}{\lambda_r}
\newcommand{\paramtrans}{\lambda_t}
\newcommand{\probmix}{p_{\mathrm{mix}}}

\begin{theorem}
 \label{thm:mix-bound}
 For a protocol $P$, assume randomization intervals are i.i.d.\ $\mathrm{Exp}(\lambda_r)$ and transmission intervals are i.i.d.\ $\mathrm{Exp}(\lambda_t)$. For two location-indistinguishable devices, let $p_{\mathrm{mix}}$ be the probability that their identifiers have mixed after the first mixing opportunity within duration $T$. Then, with $B=\lambda_r+\lambda_t$,
\begin{equation*}
 \textstyle
 p_{\mathrm{mix}}\le \lambda_r\!\left(\frac{1}{B}+\frac{\lambda_t}{B^2}\right)\!\left(1-e^{-2\lambda_r T}\right)
 \le \lambda_r\!\left(\frac{1}{B}+\frac{\lambda_t}{B^2} \right)
 =: p_{\mathrm{mix},1}^{\mathrm{ub}}(P).
\end{equation*}

\end{theorem}

\begin{theorem}
\label{thm:mix-lower_bound_multi}
Let $\probmix(T)$ be the probability that two location-indistinguishable devices mix for protocol $P$ within duration $T$, allowing multiple randomizations during $T$. Let $p_{\mathrm{mix},1}^{\mathrm{ub}}(P)$ denote our single-opportunity upper bound (Theorem~\ref{thm:mix-bound}). Then
\[
\probmix(T) \le 1 - e^{-2\lambda_r T \cdot p_{\mathrm{mix},1}^{\mathrm{ub}}(P)}
\;=:\; p_{\mathrm{mix}}^{\mathrm{ub}}(T,P).
\]
\end{theorem}
In a multi-protocol setting, perfect mixing requires \emph{every} monitored protocol to mix within the same duration $T$. Let $E_P(T)$ denote the event that protocol $P$ mixes within $T$. For a set of monitored protocols $\widehat{P}$, perfect mixing is the event $\bigcap_{P\in\widehat{P}} E_P(T)$. Under our cross-protocol independence assumption,
\begin{align}
\label{eqn:perfect_mixing}
P_{\text{mix}}(T)
= \Pr\!\left[\bigcap_{P \in \widehat{P}} E_P(T)\right]
= \prod_{P \in \widehat{P}} \Pr[E_P(T)]
\le \prod_{P \in \widehat{P}} p_{\mathrm{mix}}^{\mathrm{ub}}(T,P).
\end{align}

 \begin{table}[t]
 \centering
 \small
 \setlength{\tabcolsep}{5pt}
 \renewcommand{\arraystretch}{1.05}
 \begin{tabular}{@{}rccc@{}}
 \toprule
 $T$ & $p_{\mathrm{mix}}^{\mathrm{ub}}(T,P_A)$ & $p_{\mathrm{mix}}^{\mathrm{ub}}(T,P_B)$ & $\prod_{P\in\{P_A,P_B\}} p_{\mathrm{mix}}^{\mathrm{ub}}(T,P)$ \\
  \midrule
  $60$\,s   & $0.58$  & $7.7{\times}10^{-3}$ & $4.5{\times}10^{-3}$ \\
  $5$\,min  & $0.99$  & $0.038$              & $0.0376$ \\
  $10$\,min & $1.00$  & $0.075$              & $0.075$ \\
  $1$\,h    & $1.00$  & $0.373$              & $0.373$ \\
  \bottomrule
  \end{tabular}
  \caption{Examples for Thms~\ref{thm:mix-bound}--\ref{thm:mix-lower_bound_multi}. We use $P_A: \lambda_t^{(A)} = 0.05$, $\lambda_r^{(A)} = 1/60$ and $P_B$: $\lambda_t^{(B)} = 0.0833$, $\lambda_r^{(B)} = 1/600$ (rates in s$^{-1}$).} 

\label{tab:mix_numeric}
\end{table}

Table~\ref{tab:mix_numeric} illustrates the bottleneck effect: if one protocol mixes slowly, it dominates the overall probability of perfect mixing. Moreover, Eq.~\ref{eqn:perfect_mixing} shows that, under independent protocol behavior, the best achievable perfect-mixing probability shrinks multiplicatively as the adversary monitors more protocols. Thus, per-protocol privacy guarantees do not compose; multi-protocol devices require joint privacy analysis. We defer proofs to Appendix~\ref{app:bounds}.

\section{Real-Device Validation}
\label{sec:realdevice}

We validate two claims on real commodity devices: \name links identifiers across protocols, and it fails only when all observed protocols satisfy the perfect mixing conditions. 

\subsection{Experimental Setup}
We construct a small-scale mix-zone experiment with 12 commodity smartphones in a controlled testbed. We restrict ourselves to the LTE and BLE protocols. We exclude WiFi because its MAC rotation rate approaches its transmission rate (Section~\ref{section:characterization}), leaving too few stable intervals for controlled validation; Section~\ref{sec:evaluation} evaluates WiFi in simulation.

To ensure LTE rotations happen, we use two eNodeBs that trigger LTE handovers during the observation window. BLE rotations occur naturally.

We construct a passive adversary that can receive BLE MAC addresses and C-RNTI values of all devices. Before devices enter the mix-zone, they are well-separated in space, thus enabling the adversary to pair BLE MAC and C-RNTI values. Similarly, after devices leave the mix-zone, the adversary can pair BLE MAC and C-RNTI values. We provide these pairings as anchors to \name.

Devices stay inside the mix-zone for 40 minutes. While they are there, they are location-indistinguishable. That is, they are so close that received identifiers could originate from any of the devices.

\name's task is to connect devices (represented by a MAC, C-RNTI anchor) entering the mix zone, to devices (i.e., anchor) leaving the mix zone. We count a link as correct if the recovered mapping is unique and valid (i.e., this is really the same device). To assess correctness, we collect ground-truth data separate from the adversary's input: eNodeB logs map C-RNTIs to devices, and isolated BLE recordings label each device's MACs.

\subsection{Results}

\begin{figure}[!tp]
      \centering
      \includegraphics[]{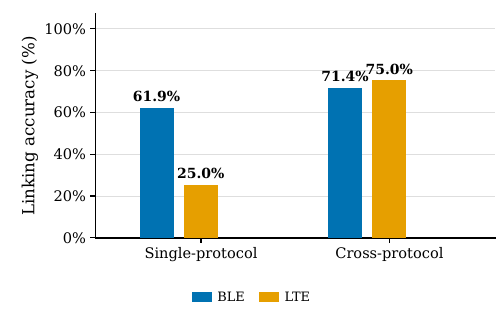}
      \caption{Accuracy of identifier mappings under single-protocol and cross-protocol observation for 12 devices in a mix-zone.}
      \Description{Comparing single vs multi-protocol observation}
      \label{fig:real_devices}
  \end{figure}

\noindent\textbf{Cross-protocol observation improves linking accuracy.} Figure~\ref{fig:real_devices} compares identifier-linking accuracy for 12 devices under single-protocol and cross-protocol observation. Cross-protocol evidence raises BLE accuracy from 61.9\% to 71.4\% and LTE accuracy from 25.0\% to 75.0\%. LTE benefits most: under single-protocol observation, when two devices experience simultaneous inter-eNodeB handovers, their LTE rotations are ambiguous. BLE singleton chains help break these ties. Conversely, when BLE rotations are ambiguous and LTE handovers are not simultaneous, stable LTE identifiers bridge the BLE rotations. These results illustrate \name's core mechanism: stability in one protocol can defeat rotation in another.

\smallskip
\noindent\textbf{\rev{\name's success or failure is consistent with perfect mixing.}} To understand where \name fails and where it succeeds, we take 4 commodity devices and repeat the same experiment. Except this time, we force \emph{all four} of the devices to do simultaneous LTE handover to create a \emph{four-way} ambiguity for each LTE rotation. For BLE, we can construct a unique chain for two out of the four devices just by following the intra-links. \name is able to use the BLE singleton chains in these two devices to resolve LTE's four-way ambiguity introduced by simultaneous inter-eNodeB handover. In this controlled experiment, \name fails to resolve the LTE rotations and reconstruct a unique chain only when
both LTE and BLE simultaneously satisfy the mixing condition. \rev{This behavior is consistent with the boundary case of Theorem~\ref{thm:mix-lower_bound_multi}.}

\section{Large-Scale Evaluation}
\label{sec:sim_eval}
Section~\ref{sec:realdevice} validated \name on real devices; in this section we assess its population-scale privacy impact. Measuring real users at this scale, however, requires the passive surveillance we warn against. Therefore, we \emph{simulate} users in a dense urban environment with transmission and rotation parameters derived from our commodity-device measurements (Section~\ref{section:characterization}) to evaluate \name across adversary capabilities, protocol parameters, and environmental conditions.

\subsection{Simulation Setup}
\label{sec:simulation_setup}

Our simulation emulates a \emph{passive} mass-surveillance deployment over a dense urban region of 3.87 $\mathrm{km}^2$. We simulate pedestrian mobility using SuMO~\cite{SUMO2018} with the Monaco scenario configuration~\cite{monaco_sumo} and run the simulation for a total duration of 2 hours. Each simulated pedestrian carries a multi-protocol device that emits BLE, WiFi, and LTE signals. For each protocol, the simulator generates (i) time-stamped transmissions and (ii) identifier rotations, based on empirical measurements (Section~\ref{section:characterization}).

An adversary deploys protocol-specific sniffers under a coverage budget. Each sniffer records the temporary identifier when user is within the sniffer reception range  \rev{together with a distance estimate. Estimates include protocol-specific localization errors.}
The placement and coverage of sniffers depends upon the adversary's budget and strategy. We consider four deployment strategies: 
\begin{enumerate}
      \item \textbf{RAND}: The adversary places a fixed number of sniffers randomly over the target region.
      \item \textbf{SPOT}: The adversary uses a spatial prior and places sniffers in overlap regions of adjacent eNodeBs, motivated by our observation that deterministic C-RNTI linking is not possible at handover; and thus requires observations.
      \item \textbf{PATCH}: The adversary provides full coverage only within a contiguous subregion of the target area and leaves the remainder unmonitored.
      \item \textbf{MOB}: The adversary uses a small number of compromised user devices as mobile sniffers, passively recording locally received transmissions. This models supply-chain threats in which malicious firmware, third-party SDKs, or bloatware turn ordinary mobile devices into sniffers.
  \end{enumerate}

\noindent\textbf{Privacy Metric.} To enable direct comparisons, we normalize all privacy metrics $\textsf{LP}_{\mathcal{D}}$ by the device's \emph{total time within the simulation}. \rev{Under full coverage, this denominator coincides with the observation duration in Definition~\ref{def:privacy}. Under partial coverage, time outside sniffer coverage remains in the denominator and therefore lowers $\textsf{LP}_{\mathcal{D}}$, unless \name links the observations before and after the gap into the same reconstructed trace.}

\begin{table}[t]
	\caption{Randomization (RI) and transmission (TI) interval values from actual devices and distributions used in our simulation, $\mathcal{E}(s)$ is exponential with scale $s$, $\mathcal{U}(u)$ is the uniform distribution on $[a,b]$. \textbf{NC:} No change observed. }
	\centering
	\small
	\scriptsize
	\begin{tabular}{@{}lllllllll}
		\toprule
		\multicolumn{1}{c}{} & & \multicolumn{2}{c}{Real Devices }  & & \multicolumn{2}{c}{Simulations}  \\
		\cmidrule{3-4}   \cmidrule{6-7} 
		Protocol & \textbf{Mode} & \textbf{TI (s)} & \textbf{RI (s)}  & & \textbf{TI (s)}  & \textbf{RI (s)}   \\
		\midrule			
		LTE      	&  Active & 0-5  &   NC           &&      &      \\
                  	&  Inactive & 0-55  &  185-593         &&   $\mathcal{E}(3)$  & $\mathcal{E}(420)$    \\
       \midrule
		WiFi   	    &  Active Conn. & 0-0.3  &  NC         &&     &   \\
            	    &  Inactive Conn. & 0-54  &  NC          &&    &     \\
            	    &  Inactive Disconn. & 0-60  &  0-60        &&   $\mathcal{U}(0,60)$  & $\mathcal{U}(0,60)$    \\
        \midrule
		Bluetooth   &  Active & 0-35  &  420-1200          &&   $\mathcal{E}(5)$  & $\mathcal{U}(420,900)$     \\
		\bottomrule
	\end{tabular}
	\label{tab:RI_TI}
\end{table}

\vspace{2pt}
\noindent\textbf{Parameter selection.}
We cap pedestrian speed at \(1.6\)~m/s, a conservative upper bound for walking. We assume inactive/background device behavior, which produces fewer transmissions and therefore avoids overestimating the attacker's ability to track users. 

\noindent\textbf{Sniffer Radius:} Sniffer reception radii are 30 \,m (BLE), 50 \,m (WiFi), and 100 \,m (LTE).

\rev{\noindent\textbf{Localization error model}: 
We evaluate \name in two localization-error models, the bounded error model (baseline) and the heavy tailed error model. Let $d$ denote the true device--sniffer range and $\dhat$ the range estimate given to \name.}

\rev{\emph{Bounded baseline.}
For each observation of protocol $p$, we draw
\[
    e_p \sim \mathcal{U}[-b_p,b_p]
    \qquad\text{and set}\qquad
    \dhat=d+e_p,
\]
where $b_p$ is the protocol-specific bound on the absolute additive error $e_p$. \name sets $\epsilon_p=b_p$ in Equation~\ref{eqn:mob}. Thus, every generated baseline error lies within the corresponding matching tolerances. Based on the error measurements reported in prior work~\cite{nussbaummuller_ble,Bao2022,ltrack}, we set
$\epsilon_{\mathrm{BLE}}=1.5$\,m,
$\epsilon_{\mathrm{WiFi}}=5$\,m, and
$\epsilon_{\mathrm{LTE}}=10$\,m as the baseline matching error tolerances.}
\rev{For BLE, the $1.5$\,m error tolerance rounds up the largest RMSE of $1.47$\,m reported by
Nu{\ss}baumm{\"u}ller et al.~\cite{nussbaummuller_ble}, while the WiFi value rounds up the largest reported MAE of
$4.62$\,m. We choose the LTE value conservatively for commercial phones: it exceeds the reported $4.67$--$7.24$\,m 90th-percentile errors and is close to the $10.47$\,m error reported for USRP B210.}

\rev{\emph{Heavy-tail error.} The bounded baseline assumes that every localization error lies within a fixed interval. Real-world distance errors are more variable: multipath, obstruction, and measurement noise can produce occasional large outliers, and the absolute error often increases with the true distance. We therefore model the estimated distance $\dhat$ under protocol $p$ as
\[
    \dhat=d\eta_p,
    \qquad
    \eta_p=\exp(\sigma_p Z),
    \qquad
    Z\sim\mathcal{N}(0,1),
\]
where $d$ is the true distance. Thus, $\eta_p$ is log-normal with median one and a right-skewed tail. Values $\eta_p<1$ underestimate the distance, whereas values $\eta_p>1$ overestimate it. We derived values of the $\sigma_p$ parameters for each protocol based prior work (see Section~\ref{sec:key-findings}): $\sigma_{\mathrm{BLE}}=0.58$, $\sigma_{\mathrm{WiFi}}=0.326$, and $\sigma_{\mathrm{LTE}}=0.061$. The details of the
error expression derivations are deferred to Appendix~\ref{app:sigma-derivation}. We use this unbounded, distance-scaled model to evaluate \name under heavy-tailed, protocol-specific localization noise}.

\noindent\textbf{Transmission and Randomization Interval.} The transmission intervals we measured for real devices are heavy-tailed. We approximate them with exponential distributions and fit per-protocol means via MLE. We conservatively round up to $5\,\textrm{s}$ (BLE) and $3\,\textrm{s}$ (LTE). WiFi is modeled as an inactive, disconnected device sending one probe per minute with fresh MAC. BLE identifier lifetimes are bounded and periodic, modeled with a uniform distribution fitted to our measurements. LTE rotations are event-driven and modeled two ways: an exponential process with 7-minute mean (rotation irrespective of mobility), or handover-triggered based on OpenCellID-derived eNodeB locations. See Table~\ref{tab:RI_TI} for a summary.

\subsection{Evaluation}
\label{sec:evaluation}
\begin{figure}[t]
    \centering

    \includegraphics[]{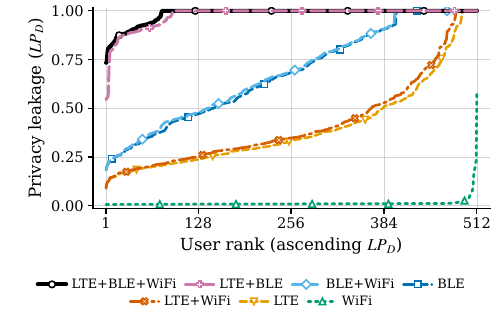}
    \caption{Sorted per-user privacy leakage under full coverage for 512 users in multi-protocol setting}
    \label{fig:privacy_leakage_q1}
    \Description{512 user full coverage}
 
\end{figure}

\begin{figure}[t]
    \centering
     \includegraphics[]{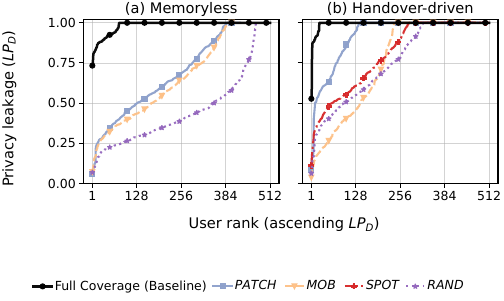}
    \caption{Privacy leakage under partial coverage with budgeted deployments. Left: LTE identifier changes follow a memoryless (exponential) model. Right: LTE identifier changes are handover-driven, creating spatially correlated regions that enable handover-aware placement}
    \label{fig:privacy_leakage_q3a}
    \Description{partial coverages}
\end{figure}

\begin{figure}[t]
     \centering
     \includegraphics[]{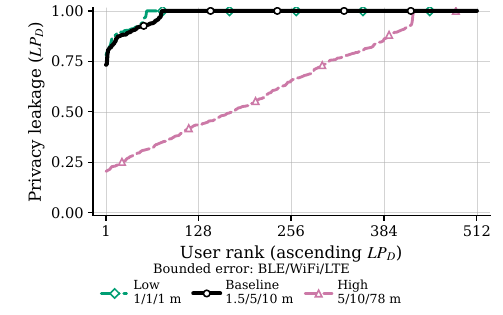}
     \caption{Privacy Leakage under different bounded localization errors}
     \label{fig:privacy_leakage_q3c}
     \Description{bounded localization error}

\end{figure}

\begin{figure}[t]
    \centering
    \includegraphics[]{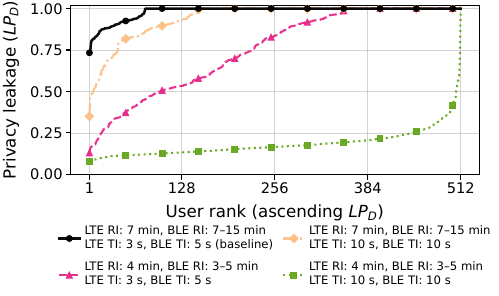}
    \caption{Privacy Leakage under different randomization and transmission interval}
    \label{fig:privacy_leakage_q2}
    \Description{Different RI and TI}
   
\end{figure}

\begin{figure*}[t]
    \centering

    \begin{subfigure}[t]{0.315\textwidth}
        \centering
        \includegraphics[
            width=\linewidth,
            keepaspectratio
        ]{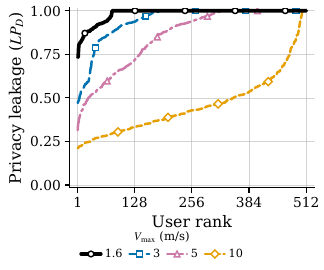}
        \caption{Varying maximum user velocity.}
        \label{fig:q4_mobility}
    \end{subfigure}
    \hfill
    \begin{subfigure}[t]{0.315\textwidth}
        \centering
        \includegraphics[
            width=\linewidth,
            keepaspectratio
        ]{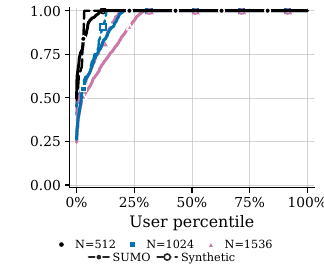}
        \caption{Varying user density.}
        \label{fig:q4_density}
    \end{subfigure}
    \hfill
    \begin{subfigure}[t]{0.315\textwidth}
        \centering
        \includegraphics[
            width=\linewidth,
            keepaspectratio
        ]{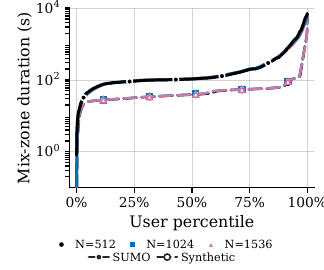}
        \caption{Mix-zone duration.}
        \label{fig:q4_mix}
    \end{subfigure}

    \caption{Effect of mobility and user density on tracking.
    For each configuration, we independently sort users by
    privacy leakage. (a) Privacy leakage for varying maximum
    user velocity in SUMO. (b) Privacy leakage for varying user
    density in SUMO and synthetic simulations. (c) Mix-zone
    duration for varying user density in SUMO and synthetic
    simulations.}
    \Description{Three side-by-side plots. The left plot compares
    privacy leakage at maximum user velocities of 1.6, 3, 5, and
    10 meters per second. The middle plot compares privacy leakage
    at different user densities in SUMO and synthetic simulations.
    The right plot compares mix-zone duration for the same user
    densities and simulation settings.}
    \label{fig:privacy_leakage_merged}
\end{figure*}

\begin{figure}[tbp]
    \centering
    \includegraphics[]{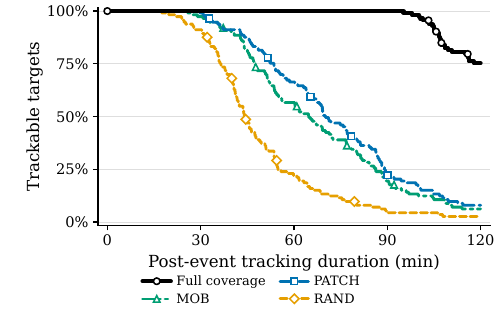}
    \Description{Targeted scenario}
    \caption{Tracking duration of users after attending the protest event under different adversary strategies}
    \label{fig:privacy_leakage_targeted}
\end{figure}

\begin{figure}[tbp]
    \centering
    \includegraphics[]{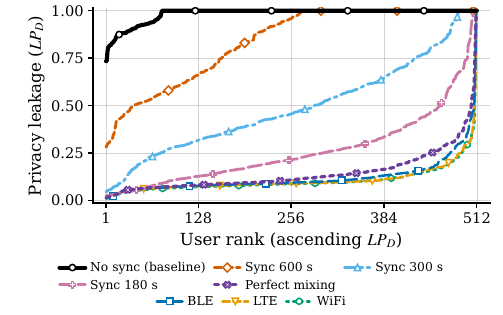}
    \caption{Privacy leakage under different countermeasure techniques versus single protocol baselines (BLE, LTE, WiFi)}
    \Description{countermeasure}
    \label{fig:privacy_leakage_q5}
\end{figure}

We evaluate the tracking mechanism by measuring privacy leakage for each user as defined in Definition~\ref{def:privacy}. Since we simulate user movements and sniffer observations, ground truth data is fully accessible for calculating these metrics. Figures~\ref{fig:privacy_leakage_q1}, \ref{fig:privacy_leakage_q3a}--\ref{fig:privacy_leakage_q5} \rev{show the sorted per-user privacy-leakage. We independently sort users by $\textsf{LP}_{\mathcal{D}}$ in ascending order. The x-axis shows the resulting user rank, while the y-axis shows the $\textsf{LP}_{\mathcal{D}}$ of the user at that rank.} A larger area under the curve (AUC) indicates greater privacy loss. In our evaluation, we aim to answer the following questions:

\smallskip
\textbf{Q1. How effectively can an adversary conduct physical tracking of users through cross-linking?}

\begin{figure}[t]
    \centering

    \includegraphics[]{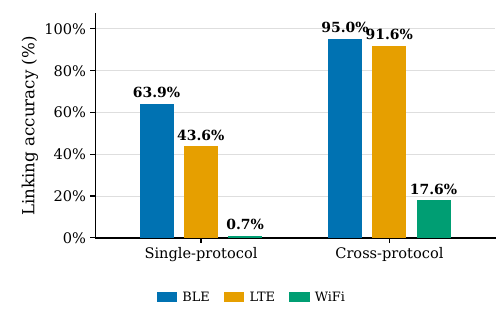}
    \caption{Accuracy of identifier mappings for 512 simulated users under single- and cross-protocol observation.
    }
    \label{fig:privacy_leakage_q1_acc}
    \Description{accuracy of identifier linkings}
 
\end{figure}

We first evaluate an idealized full-coverage baseline in which sniffers are placed optimally using hexagonal tiling~\cite{Alam2006Hexagonal} such that any transmission by the devices can be received by at least one sniffer with high probability. In the single-protocol baselines, we run the same tracing pipeline but disable inter-protocol edges, i.e., we follow only intra-protocol links. Due to WiFi’s rapid identifier rotation, users are difficult to track when relying solely on WiFi.

\rev{Figure~\ref{fig:privacy_leakage_q1} shows the sorted per-user privacy-leakage ($\textsf{LP}_{\mathcal{D}}$) curve for 512 pedestrians.} With cross-protocol linkage (LTE--BLE--WiFi), 83.39\% of users have $\textsf{LP}_{\mathcal{D}}=1$, and 90.03\% have $\textsf{LP}_{\mathcal{D}}\ge 0.9$. In contrast, the single-protocol baselines recover full traces ($\textsf{LP}_{\mathcal{D}}=1$) for only 21.87\% of users in BLE and 4.1\% in LTE; no user is fully reconstructed using WiFi alone. These results show that combining protocols substantially increases traceability. Overall, LTE--BLE--WiFi yields the highest privacy leakage. LTE--BLE alone reconstructs 82.61\% of users ($\textsf{LP}_{\mathcal{D}}=1$), nearly matching LTE–BLE–WiFi: WiFi adds little because its per-probe randomization.

To better understand this gain, Figure~\ref{fig:privacy_leakage_q1_acc} reports the accuracy of identifier rotations, i.e., the fraction of identifier rotations that are singleton links (no ambiguity) and correspond to the correct same-user out of the total identifier rotation. Figure~\ref{fig:privacy_leakage_q1_acc} shows that cross-protocol observation increases singleton-link accuracy from 63.9\% to 95.0\% for BLE, 43.6\% to 91.6\% for LTE, and 0.7\% to 17.6\% for WiFi. WiFi's absolute accuracy remains limited because rapid identifier rotation leaves little linking opportunity to exploit. This result suggests that the privacy-leakage gains are driven by more accurate identifier linkage through cross-protocol observations resulting in longer reconstructed traces.

\smallskip
\textbf{Q2. How do different parameters of the adversary's capability, like localization error and coverage budget, impact the performance of the tracking?}

In practice, a fully passive adversary may be budget-constrained and unable to provide dense citywide coverage. Blind spots fragment mobility traces; in that case, cross-protocol linking depends not only on the number of sniffers deployed, but also on their placement strategy. We evaluate four partial-coverage deployments that represent different adversarial capabilities, as shown in Figure~\ref{fig:privacy_leakage_q3a}. For LTE rotation, we consider both the memoryless model (left) and the handover-driven rotation model (right).

In our experiments, \emph{RAND} and \emph{SPOT} use the same fixed-sniffer budget (\(N{=}198\)) to isolate the effect of placement, while \emph{MOB} uses \(N{=}30\) randomly selected mobile sniffers. Figure~\ref{fig:privacy_leakage_q3a} shows that partial coverage lowers privacy leakage because traces break when users enter blind spots; however, the deployment strategy strongly affects the remaining linkability. \emph{RAND} performs worst because many receivers land in low-traffic/low-utility areas and miss transmissions. \emph{PATCH} performs better by concentrating sniffers to achieve near-complete coverage within a contiguous subregion, producing reliable observations while users remain inside that zone. \emph{MOB} demonstrates feasibility under limited budget constraints: even with few sniffers, mobility sweeps a large spatial footprint over time and accumulates observations along long trajectories. Under the same budget (\(N{=}198\)), Figure~\ref{fig:privacy_leakage_q3a} shows \emph{SPOT} consistently outperforms \emph{RAND} by placing sniffers in eNodeB overlap (handover) regions where LTE identifiers are most likely to rotate, increasing the chance of observing the rotation event and stitching traces across it. The results also show that mobile sniffers, represented by the \emph{MOB} strategy, outperform random static deployments, indicating that even a small number of compromised smartphones can compromise the location privacy of other individuals.

\begin{table}[t]
\caption{\rev{Effect of error-tolerance choices on the effectiveness of \name for different error models in setting with full coverage and 512 users.}}
\label{tab:heavy_tail}

\begin{minipage}{\columnwidth}
\centering
\footnotesize
\setlength{\tabcolsep}{2pt}
\renewcommand{\arraystretch}{1.10}

\begin{tabular}{@{}
    >{\raggedright\arraybackslash}p{0.45\linewidth}
    >{\centering\arraybackslash}p{0.15\linewidth}
    >{\centering\arraybackslash}p{0.14\linewidth}
    >{\centering\arraybackslash}p{0.17\linewidth}
@{}}
\toprule
Error model &
$\epsilon_p$ (m) &
$LP_D =1$ &
$LP_D \geq 0.9$ \\
\midrule

Bounded (baseline)
    & 1.5/5/10
    & 83.39\%
    & 90.03\% \\

Heavy-tail (baseline error tolerance)
    & 1.5/5/10
    & 0.90\%
    & 1.10\% \\

Heavy-tail (increased error tolerance)
    & $15/20/20$
    & 60.93\%
    & 73.04\% \\

\bottomrule
\end{tabular}
\end{minipage}
\end{table}

\rev{Table~\ref{tab:heavy_tail} evaluates the full-coverage, 512-user setting under the bounded and heavy-tailed error model defined in Section~\ref{sec:simulation_setup}. With bounded localization error and matching baseline, \name reconstructs full traces for $83.39\%$ of users and achieves $LP_D \geq 0.9$ for $90.03\%$ of users.  Applying the same error tolerances under heavy-tailed error reduces these values to $0.90\%$ and $1.10\%$, respectively (row~2). Heavy-tailed range errors can exceed $\epsilon_p$, causing Equation~\ref{eqn:mob} to reject true links during candidate set construction. Refinement cannot recover these links because it only removes candidates.
However, the adversary can increase $\epsilon_p$ conservatively to retain more true links, at the cost of admitting more false candidates. Cross-protocol refinement removes enough of the additional false candidates to reconstruct full traces for $60.93\%$ of users, with $LP_D \geq 0.9$ for $73.04\%$ of the users (row~3). The remaining gap from row~1 reflects the additional ambiguity. Thus, higher localization uncertainty reduces privacy leakage but does not eliminate the attack. \name can gracefully deal with heavy-tail errors by admitting larger initial candidate sets and using cross-protocol consistency to prune the resulting false links.}

\rev{Finally we directly evaluate the effects of increasing the} per-protocol localization error (Figure~\ref{fig:privacy_leakage_q3c}). Privacy leakage increases as localization becomes more precise. This effect will likely grow with improved receivers: wider bandwidths, larger MIMO arrays, and \rev{angle of arrival} support reduce localization error, shorten the mix-zone duration in Theorem~\ref{thm:mix-lower_bound_multi}, and lower the probability of perfect mixing. If rotation behavior remains unchanged, better receivers will further weaken location privacy. 

\smallskip
\textbf{Q3. How do different parameters of wireless communication protocols, such as randomization interval and transmission interval, affect the success of tracking?}

Figure~\ref{fig:privacy_leakage_q2} shows the sensitivity of tracking performance to the transmission interval (TI) and randomization interval (RI). Holding other parameters fixed, privacy leakage decreases as TI increases (fewer observable transmissions) and as RI decreases (more frequent identifier rotation). Our baseline is intentionally conservative: we parameterize TI/RI using inactive/background device behavior and choose transmission-interval distributions that do not overestimate the attacker (Figures~\ref{fig:ti_lte_combined} and ~\ref{fig:ble_ti_ri_combined}). Consequently, in real deployments where devices transmit more frequently, leakage may be higher for comparable randomization behavior. 

\smallskip
\textbf{Q4. How do other environmental factors, such as the density of users and the velocity of movement, affect the surveillance performance?} 

\label{para:q4_mobility}
Figure~\ref{fig:privacy_leakage_merged} (left) shows that privacy leakage decreases as users move faster. This reflects a limitation of our mobility model: Equation~\ref{eqn:mob} admits feasible but physically unrealizable paths, so the reported leakage is conservative; a more accurate mobility model would likely track better.

Figure~\ref{fig:privacy_leakage_merged} (middle) shows how privacy leakage varies with user density. Privacy leakage decreases with user density. We conjectured that this occurs because users have more opportunities to encounter other users, and thus confuse the tracking algorithms when they rotate identifiers. To confirm, we re-ran simulations with synthetic random-walk trajectories. We tuned the number of edges in the random walk to force a decrease in user interaction with respect to SuMO (we ended up with 80\%). Figure~\ref{fig:privacy_leakage_merged} (right) shows the duration for which users are so close to another they are indistinguishable to sniffers i.e., the mix-zone duration. edges  Figure~\ref{fig:privacy_leakage_merged} (right) confirms the drop in interaction time. Figure~\ref{fig:privacy_leakage_merged} (middle) shows that when users interact less often (as for synthetic data) tracking improves. 

\smallskip
\noindent\textbf{Example Protest Scenario:}
We instantiate the risk described in Section~\ref{sec:motivation}: a device observed at a protest becomes a target, and the adversary attempts to track it after it leaves the event. In our simulation, we designate an area as protest site and identify 113 devices during the protest. For each protester's device, we measure how long the adversary can continuously track the device after it exits the event region. Figure~\ref{fig:privacy_leakage_targeted} plots, for each duration \(t\), the fraction of targets that remain trackable for at least \(t\) minutes. We show that even budget-limited deployments sustain long post-event tracking: 75 targets (66\%) remain trackable for at least one hour with \emph{PATCH}, 64 (57\%) with \emph{MOB}, and 25 (22\%) with \emph{RAND}. The median tracking durations are 70, 65, and 44 minutes, respectively, long enough to associate many protest attendees with later visits to sensitive locations such as homes or workplaces.

\smallskip
\noindent
\rev{\textbf{Sniffer deployment and cost.}
\label{par:sniffer_cost}
In \emph{MOB}, the adversary uses 30 compromised smartphones as mobile sniffers. This strategy may incur little or no additional hardware cost if the adversary can compromise existing devices. For all other settings, we assume a per-sniffer hardware cost of \$10 for BLE (nRF52840), \$40 for WiFi (ALFA AWUS036ACM), and \$150 for LTE (HackRF One). The resulting deployment cost therefore depends on the number and type of sniffers required by each adversarial strategy. In the Monaco setting~\cite{monaco_sumo} using SuMO~\cite{SUMO2018} \emph{RAND} and \emph{SPOT} each use 198 sniffers per protocol, resulting in a total hardware cost of approximately \$40k.  In contrast, the full-coverage setting requires 1778 BLE, 670 WiFi, and 185 LTE sniffers, increasing the total hardware cost to approximately \$72.3k. The large number of BLE receivers is mainly due to BLE's shorter reception range, which requires a denser deployment to achieve full coverage. These estimates reflect current hardware prices; as commodity receiver hardware becomes less expensive, the cost of deploying such large-scale adversarial sensing infrastructures may decrease further.}

\smallskip
\noindent\textbf{Countermeasure:}
Cross-protocol tracking arises because temporary identifiers rotate independently. We therefore evaluate two synchronization-based defenses in Figure~\ref{fig:privacy_leakage_q5}. \emph{Perfect mixing} forces co-located devices to rotate identifiers together, eliminating cross-protocol bridges and limiting tracking to a single identifier lifetime; however, it requires costly coordination across devices. Figure~\ref{fig:privacy_leakage_q5} shows that the privacy leakage in \emph{Perfect Mixing} approaches that of the naive
BLE, WiFi, and LTE-only identifier-lifetime baselines. These naive baselines do not link identifiers across rotations and therefore limit each reconstructed trace to one identifier lifetime. \emph{Local synchronization} instead rotates all identifiers on the same device simultaneously, without coordinating with other users. Its effectiveness depends on the rotation period: with 10-minute synchronization, 254 of 512 users remain fully trackable, whereas reducing the period to 3 minutes lowers this number to 54 by shortening identifier lifetimes and increasing incidental mixing among nearby devices.

These results underscore the difficulty of achieving robust privacy through synchronization alone. Per-transmission identifiers~\cite{SingeleeP06} could approximate perfect mixing, but would incur substantial overhead and limit usability. Designing countermeasures offering strong privacy while preserving utility thus remains an open problem.

\section{Conclusion}

This paper asks whether identifier randomization remains effective when a smartphone is observed simultaneously for multiple protocols. We show that it does not. Even when each protocol leaks only temporary identifiers and the adversary is fully passive, unsynchronized rotations across protocols create linking opportunities that restore continuity across time and space. \name exploits this system-level composition failure to reconstruct long mobility traces under noisy localization and asynchronous transmissions. We further show that this threat does not require dense, citywide infrastructure: even with a small number of mobile sniffers, a limited-budget adversary can gather enough observations to construct mobility traces of the wider population.

\section{Acknowledgments}

We thank Sylvain Chatel for feedback on the paper at various stages of the submission process. We are grateful for the constructive feedback from the anonymous reviewers and for the opportunity to conduct the revision process. Aneet Kumar Dutta carried out this work as a member of the Saarbrucken Graduate School of Computer Science.

\bibliographystyle{ACM-Reference-Format}

\bibliography{references}

@article{BeresfordS03,
  author       = {Alastair R. Beresford and
                  Frank Stajano},
  title        = {{Location Privacy in Pervasive Computing}},
  journal      = {{IEEE} Pervasive Comput.},
  volume_       = {2},
  number_       = {1},
  pages_        = {46--55},
  year         = {2003},
  url_          = {https://doi.org/10.1109/MPRV.2003.1186725},
  doi_          = {10.1109/MPRV.2003.1186725},
  timestamp_    = {Tue, 31 Mar 2020 18:15:52 +0200},
  biburl_       = {https://dblp.org/rec/journals/pervasive/BeresfordS03.bib},
  bibsource_    = {dblp computer science bibliography, https://dblp.org}
}

@misc{3gpp136300_ho,
 author = {{3GPP}},
 title = {{LTE; Evolved Universal Terrestrial Radio Access (E-UTRA) and Evolved Universal Terrestrial Radio Access Network (E-UTRAN); Overall Description}},
 howpublished = {3GPP TS 36.300 v17.9.0, Release 17},
 booktitle_ = {3GPP TS 36.300 version version 17.9.0 Release 17},
 year = {2024}, 
 url_ = "https://www.etsi.org/deliver/etsi_ts/136300_136399/136300/17.09.00_60/ts_136300v170900p.pdf"
}

@article{nussbaummuller_ble,
  author  = {Barbara Nu{\ss}baumm{\"u}ller and
             Bernhard Etzlinger and
             Karin Anna Hummel},
  title   = {{BLE}-Based Contact Tracing:
             Characterization of Distance Estimation Errors
             and Mitigation Options},
  journal = {IEEE Pervasive Comput.},
  volume_  = {22},
  number_  = {4},
  pages_   = {7--14},
  year    = {2023},
  doi_     = {10.1109/MPRV.2023.3323747}
}

@INPROCEEDINGS{BeresfordS04,
  author={Alastair R. Beresford and Frank Stajano},
  booktitle={IEEE PerCom Workshops}, 
  title={{Mix Zones: User Privacy in Location-Aware Services}}, 
  year={2004},
  volume_={},
  number_={},
  pages_={127-131},
  keywords_={Middleware;Computational complexity;Feedback;Data privacy;Hospitals;Protection;Shape measurement;Laboratories;Mathematical model;Tracking},
  doi_={10.1109/PERCOMW.2004.1276918}}

@article{GruteserG05,
  author       = {Marco Gruteser and
                  Dirk Grunwald},
  title        = {{Enhancing Location Privacy in Wireless {LAN} Through Disposable Interface Identifiers: {A} Quantitative Analysis}},
  journal      = {Mob. Networks Appl.},
  volume_       = {10},
  number_       = {3},
  pages_        = {315--325},
  year         = {2005},
  url_= {https://doi.org/10.1007/s11036-005-6425-1},
  doi_= {10.1007/S11036-005-6425-1},
  timestamp_= {Thu, 24 Sep 2020 14:11:18 +0200},
  biburl_       = {https://dblp.org/rec/journals/monet/GruteserG05.bib},
  bibsource_= {dblp computer science bibliography, https://dblp.org}
}

@inproceedings{SingeleeP06,
  author       = {Dave Singel{\'{e}}e and
                  Bart Preneel},
  editor_       = {Radha Poovendran and
                  Ari Juels},
  title        = {{Location Privacy in Wireless Personal Area Networks}},
  booktitle_    = {Proceedings of the 2006 {ACM} Workshop on Wireless Security, Los Angeles,
                  California, USA, September 29, 2006},
  pages_        = {11--18},
  booktitle    = {{ACM} Workshop on Wireless Security},
  year         = {2006},
  url_= {https://doi.org/10.1145/1161289.1161292},
  doi_= {10.1145/1161289.1161292},
  timestamp_= {Tue, 06 Nov 2018 11:06:57 +0100},
  biburl_      = {https://dblp.org/rec/conf/ws/SingeleeP06.bib},
  bibsource_= {dblp computer science bibliography, https://dblp.org}
}

@inproceedings {ltrack,
author = {Martin Kotuliak and Simon Erni and Patrick Leu and Marc R{\"o}schlin and Srdjan Capkun},
title = {{LTrack}: Stealthy Tracking of Mobile Phones in {LTE}},
booktitle = {USENIX Sec},
year = {2022},
isbn_ = {978-1-939133-31-1},
address_ = {Boston, MA},
pages_ = {1291--1306},
url_ = {https://www.usenix.org/conference/usenixsecurity22/presentation/kotuliak},
publisher_ = {USENIX Association},
month_ = aug
}

@INPROCEEDINGS{ble_phy,
  author={Givehchian, Hadi and Bhaskar, Nishant and Herrera, Eliana Rodriguez and Soto, Héctor Rodrigo López and Dameff, Christian and Bharadia, Dinesh and Schulman, Aaron},
  booktitle={IEEE S\&P},
  title={Evaluating Physical-Layer BLE Location Tracking Attacks on Mobile Devices}, 
  year={2022},
  volume_={},
  number_={},
  pages_={1690-1704},
  doi_={10.1109/SP46214.2022.9833758}}

@inproceedings{trackingble,
    author = {Johannes K Becker and David Li and David Starobinski},
    title = {{Tracking Anonymized Bluetooth Devices}},
    booktitle = {PoPETs},
    year = {2019}
}

@misc{srsran,
	author = {srs{R}{A}{N}},
	title = {{Open Source SDR 4G Software Suite from Software Radio Systems (SRS)}},
	howpublished = {\url{https://github.com/srsran/srsRAN_4G}},
	year = {2023},
}

@misc{bluetoothprivacy,
	author = {M. Woolley},
	title = {{{B}luetooth {T}echnology - {P}rotecting {Y}our {P}rivacy}},
	howpublished = {\url{https://www.bluetooth.com/blog/bluetooth-technology-protecting-your-privacy/}},
	year= {2015},
}

@inproceedings{ltesniffer,
author = {Hoang, Tuan Dinh and Park, CheolJun and Son, Mincheol and Oh, Taekkyung and Bae, Sangwook and Ahn, Junho and Oh, BeomSeok and Kim, Yongdae},
title = {{LTESniffer: An Open-source LTE Downlink/Uplink Eavesdropper}},
year = {2023},
isbn_ = {9781450398596},
publisher_ = {ACM WiSec},
address_ = {New York, NY, USA},
url_ = {https://doi.org/10.1145/3558482.3590196},
doi_ = {10.1145/3558482.3590196},
booktitle = {ACM WiSec},
pages_ = {43–48},
numpages_ = {6},
location_ = {<conf-loc>, <city>Guildford</city>, <country>United Kingdom</country>, </conf-loc>},
series__ = {WiSec '23}
}

@misc{patent:inactivity_timer_nokia,
 title     = {Inactivity timer evaluation},
 note      = {Patent 20190268966},
 author    = {Martin Kollar and Yi Zhi Yao},
 year      = {2019},
 month_     = "August",
 url_       = "https://www.freepatentsonline.com/y2019/0268966.html"
}

@misc{inactivity_timer,
    author = {Simone Roma},
    title = {{Monitoring and testing in LTE networks: from experimental analysis to operational optimisation}},
    year = {2016},
   url="https://etd.adm.unipi.it/t/etd-05222016-160934"
}

@misc{3gpp32125,
  author       = {{3GPP}},
  title        = {{Telecommunication Management; Performance Management (PM); Performance Measurements Evolved Universal Terrestrial Radio Access Network (E-UTRAN)}},
  howpublished = {{3GPP TS 32.425 version 18.0.0 Release 18}},
  year         = {2024},
}

@misc{3gpp136304,
 author = {{3GPP}},
 title = {{LTE; Evolved Universal Terrestrial Radio Access (E-UTRA); User Equipment (UE) Procedures in Idle Mode}},
 howpublished = {3GPP TS 36.304 v18.1.0, Release 18},
 booktitle_ = {3GPP TS 36.331 version 17.6.0 Release 17},
 year = 2024,
month_ = 05,
 url_ = "https://www.etsi.org/deliver/etsi_ts/136300_136399/136304/18.01.00_60/ts_136304v180100p.pdf"
}

@INPROCEEDINGS{breaking_layer_two,
  author={Rupprecht, David and Kohls, Katharina and Holz, Thorsten and Pöpper, Christina},
  booktitle={IEEE S\&P}, 
  title={{Breaking LTE on Layer Two}}, 
  year={2019},
  volume_={},
  number_={},
  pages_={1121-1136},
  keywords_={Long Term Evolution;IP networks;Encryption;Media Access Protocol;LTE;Website-Fingerprinting;Identity-Mapping;Attack;DNS-Redirection;Mobile-Security},
  doi_={10.1109/SP.2019.00006}}

@INPROCEEDINGS{privacy_public_safety,
  author={Ghafghazi, Hamidreza and El Mougy, Amr and Mouftah, Hussein T.},
  booktitle={IEEE LCN Workshops}, 
  title={Enhancing the privacy of LTE-based public safety networks}, 
  year={2014},
  volume_={},
  number={},
  pages_={753-760},
  keywords_={Privacy;Long Term Evolution;Cryptography;Authentication;Mobile communication;Computer architecture},
  doi_={10.1109/LCNW.2014.6927731}}

@misc{jover2016lte,
      title={{LTE security, protocol exploits and location tracking experimentation with low-cost software radio}}, 
      author={Roger Piqueras Jover},
      year={2016},
      eprint={1607.05171},
      archivePrefix={arXiv},
      primaryClass={cs.CR}
}

@inproceedings{lte_sniffer_1,
author = {Kumar, Swarun and Hamed, Ezzeldin and Katabi, Dina and Erran Li, Li},
title = {{LTE Radio Analytics Made Easy and Accessible}},
year = {2014},
isbn_ = {9781450328364},
publisher_ = {ACM},
address_ = {New York, NY, USA},
url_ = {https://doi.org/10.1145/2619239.2626320},
doi_ = {10.1145/2619239.2626320},
abstract_ = {Despite the rapid growth of next-generation cellular networks, researchers and end-users today have limited visibility into the performance and problems of these networks. As LTE deployments move towards femto and pico cells, even operators struggle to fully understand the propagation and interference patterns affecting their service, particularly indoors. This paper introduces LTEye, the first open platform to monitor and analyze LTE radio performance at a fine temporal and spatial granularity. LTEye accesses the LTE PHY layer without requiring private user information or provider support. It provides deep insights into the PHY-layer protocols deployed in these networks. LTEye's analytics enable researchers and policy makers to uncover serious deficiencies in these networks due to inefficient spectrum utilization and inter-cell interference. In addition, LTEye extends synthetic aperture radar (SAR), widely used for radar and backscatter signals, to operate over cellular signals. This enables businesses and end-users to localize mobile users and capture the distribution of LTE performance across spatial locations in their facility. As a result, they can diagnose problems and better plan deployment of repeaters or femto cells. We implement LTEye on USRP software radios, and present empirical insights and analytics from multiple AT&amp;T and Verizon base stations in our locality.},
booktitle = {ACM SIGCOMM},
pages_ = {211–222},
numpages_ = {12},
keywords_ = {wireless, cellular, analytics, PHY, LTE},
location_ = {Chicago, Illinois, USA},
series_ = {SIGCOMM '14}
}

@conference{shaik_2016,
  title = {{Practical Attacks Against Privacy and Availability in 4G/LTE}},
  author = "Altaf Shaik and Ravishankar Borgaonkar and Jean-Pierre Seifert and N. Asokan and Valtteri Niemi",
  year = "2016",
  booktitle = "NDSS",
  url_ = "http://www.internetsociety.org/events/ndss-symposium-2016",
}

@inproceedings{adaptover,
author = {Erni, Simon and Kotuliak, Martin and Leu, Patrick and Roeschlin, Marc and Capkun, Srdjan},
title = {{AdaptOver: Adaptive Overshadowing Attacks in Cellular Networks}},
year = {2022},
isbn_ = {9781450391818},
booktitle = {ACM MobiCom},
address_ = {New York, NY, USA},
url_ = {https://doi.org/10.1145/3495243.3560525},
doi_ = {10.1145/3495243.3560525},
abstract_ = {In cellular networks, attacks on the communication link between a mobile device and the core network significantly impact privacy and availability. Up until now, fake base stations have been required to execute such attacks. Since they require a continuously high output power to attract victims, they are limited in range and can be easily detected both by operators and dedicated apps on users' smartphones.This paper introduces AdaptOver---a MITM attack system designed for cellular networks, specifically for LTE and 5G-NSA. AdaptOver allows an adversary to decode, overshadow (replace) and inject arbitrary messages over the air in either direction between the network and the mobile device. Using overshadowing, AdaptOver can cause a persistent (≥ 12h) DoS or a privacy leak by triggering a UE to transmit its persistent identifier (IMSI) in plain text. These attacks can be launched against all users within a cell or specifically target a victim based on its phone number.We implement AdaptOver using a software-defined radio and a low-cost amplification setup. We demonstrate the effects and practicality of the attacks on a live operational LTE and 5G-NSA network with a wide range of smartphones. Our experiments show that AdaptOver can launch an attack on a victim more than 3.8km away from the attacker. Given its practicability and efficiency, AdaptOver shows that existing countermeasures that are focused on fake base stations are no longer sufficient, marking a paradigm shift for designing security mechanisms in cellular networks.},
booktitle_ = {Proceedings of the 28th Annual International Conference on Mobile Computing And Networking},
pages_ = {743–755},
numpages_ = {13},
keywords_ = {cellular networks, denial of service, overshadowing, privacy},
location_ = {Sydney, NSW, Australia},
series_ = {MobiCom '22}
}

@inproceedings{wifi_mac,
author = {Vanhoef, Mathy and Matte, C\'{e}lestin and Cunche, Mathieu and Cardoso, Leonardo S. and Piessens, Frank},
title = {{Why MAC Address Randomization is not Enough: An Analysis of Wi-Fi Network Discovery Mechanisms}},
year = {2016},
isbn_ = {9781450342339},
booktitle = {ACM AsiaCCS},
address_ = {New York, NY, USA},
url_ = {https://doi.org/10.1145/2897845.2897883},
doi_ = {10.1145/2897845.2897883},
abstract_ = {We present several novel techniques to track (unassociated) mobile devices by abusing features of the Wi-Fi standard. This shows that using random MAC addresses, on its own, does not guarantee privacy. First, we show that information elements in probe requests can be used to fingerprint devices. We then combine these fingerprints with incremental sequence numbers, to create a tracking algorithm that does not rely on unique identifiers such as MAC addresses. Based on real-world datasets, we demonstrate that our algorithm can correctly track as much as 50\% of devices for at least 20 minutes. We also show that commodity Wi-Fi devices use predictable scrambler seeds. These can be used to improve the performance of our tracking algorithm. Finally, we present two attacks that reveal the real MAC address of a device, even if MAC address randomization is used. In the first one, we create fake hotspots to induce clients to connect using their real MAC address. The second technique relies on the new 802.11u standard, commonly referred to as Hotspot 2.0, where we show that Linux and Windows send Access Network Query Protocol (ANQP) requests using their real MAC address.},
booktitle_ = {Proceedings of the 11th ACM on Asia Conference on Computer and Communications Security},
pages_ = {413–424},
numpages_ = {12},
keywords_ = {wifi, wi-fi, tracking, scrambler, pseudonym, privacy, karma attack, hotspot 2.0, fingerprinting, anonymity, MAC address randomization, 802.11u, 802.11},
location_ = {Xi'an, China},
series_ = {ASIA CCS '16}
}

@inproceedings{lte_tracking_1_imsi,
author = {Dabrowski, Adrian and Pianta, Nicola and Klepp, Thomas and Mulazzani, Martin and Weippl, Edgar},
title = {{IMSI-Catch Me If You Can: IMSI-Catcher-Catchers}},
year = {2014},
isbn_ = {9781450330053},
booktitle={ACM ACSAC},
publisher_ = {ACM ACSAC},
address_ = {New York, NY, USA},
url_ = {https://doi.org/10.1145/2664243.2664272},
doi_ = {10.1145/2664243.2664272},
abstract_ = {IMSI Catchers are used in mobile networks to identify and eavesdrop on phones. When, the number of vendors increased and prices dropped, the device became available to much larger audiences. Self-made devices based on open source software are available for about US$ 1,500.In this paper, we identify and describe multiple methods of detecting artifacts in the mobile network produced by such devices. We present two independent novel implementations of an IMSI Catcher Catcher (ICC) to detect this threat against everyone's privacy. The first one employs a network of stationary (sICC) measurement units installed in a geographical area and constantly scanning all frequency bands for cell announcements and fingerprinting the cell network parameters. These rooftop-mounted devices can cover large areas. The second implementation is an app for standard consumer grade mobile phones (mICC), without the need to root or jailbreak them. Its core principle is based upon geographical network topology correlation, facilitating the ubiquitous built-in GPS receiver in today's phones and a network cell capabilities fingerprinting technique. The latter works for the vicinity of the phone by first learning the cell landscape and than matching it against the learned data. We implemented and evaluated both solutions for digital self-defense and deployed several of the stationary units for a long term field-test. Finally, we describe how to detect recently published denial of service attacks.},
booktitle_ = {Proceedings of the 30th Annual Computer Security Applications Conference},
pages_ = {246–255},
numpages_ = {10},
location_ = {New Orleans, Louisiana, USA},
series_ = {ACSAC '14}
}

@inproceedings{wifi_timing,
author = {Matte, C\'{e}lestin and Cunche, Mathieu and Rousseau, Franck and Vanhoef, Mathy},
title = {{Defeating MAC Address Randomization Through Timing Attacks}},
year = {2016},
isbn_ = {9781450342704},
publisher_ = {ACM},
address_ = {New York, NY, USA},
url_ = {https://doi.org/10.1145/2939918.2939930},
doi_ = {10.1145/2939918.2939930},
abstract_ = {MAC address randomization is a common privacy protection measure deployed in major operating systems today. It is used to prevent user-tracking with probe requests that are transmitted during IEEE 802.11 network scans. We present an attack to defeat MAC address randomization through observation of the timings of the network scans with an off-the-shelf Wi-Fi interface. This attack relies on a signature based on inter-frame arrival times of probe requests, which is used to group together frames coming from the same device although they use distinct MAC addresses. We propose several distance metrics based on timing and use them together with an incremental learning algorithm in order to group frames. We show that these signatures are consistent over time and can be used as a pseudo-identifier to track devices. Our framework is able to correctly group frames using different MAC addresses but belonging to the same device in up to 75\% of the cases. These results show that the timing of 802.11 probe frames can be abused to track individual devices and that address randomization alone is not always enough to protect users against tracking.},
booktitle = {ACM WiSec},
pages_ = {15–20},
numpages_ = {6},
keywords_ = {802.11, mac address randomization, privacy, security, tracking},
location_ = {Darmstadt, Germany},
series_ = {WiSec '16}
}

@INPROCEEDINGS{asimov,
  author={Ribeiro, Rafael Hengen and Rodrigues, Bruno Bastos and Killer, Christian and Baumann, Lenz and Franco, Muriel Figueredo and Scheid, Eder John and Stiller, Burkhard},
  booktitle={IFIP Networking}, 
  title={{ASIMOV: A Fully Passive WiFi Device Tracking}}, 
  year={2021},
  volume_={},
  number_={},
  pages_={1-3},
  keywords_={Location awareness;Privacy;Media Access Protocol;Mobile handsets;Hardware;Received signal strength indicator;Object recognition},
  doi_={10.23919/IFIPNetworking52078.2021.9472786}}

@INPROCEEDINGS{blc_linking,
  author={Ludant, Norbert and Vo-Huu, Tien D. and Narain, Sashank and Noubir, Guevara},
  booktitle={IEEE S\&P}, 
  title={{Linking Bluetooth LE \& Classic and Implications for Privacy-Preserving Bluetooth-Based Protocols}}, 
  year={2021},
  volume={},
  number={},
  pages_={1318-1331},
  keywords_={Meters;Privacy;Bluetooth;Protocols;Mobile handsets;Mobile applications;Internet;Bluetooth;BLE;Privacy;Linkage attacks;Contact Tracing;Exposure Notification;Apple Find My},
  doi_={10.1109/SP40001.2021.00102}}

@INPROCEEDINGS{reza_loc,
  author={Shokri, Reza and Theodorakopoulos, George and Le Boudec, Jean-Yves and Hubaux, Jean-Pierre},
  booktitle={IEEE S\&P}, 
  title={{Quantifying Location Privacy}}, 
  year={2011},
  volume={},
  number={},
  pages_={247-262},
  keywords_={Privacy;Measurement;Accuracy;Mobile communication;Gold;Random variables;Data privacy;Location Privacy;Evaluation Framework;Location Traces;Quantifying Metric;Location-Privacy Meter},
  doi_={10.1109/SP.2011.18}}

@inproceedings{Alam2006Hexagonal,
author = {Alam, S. M. Nazrul and Haas, Zygmunt J.},
title = {{Coverage and Connectivity in Three-Dimensional Networks}},
year = {2006},
isbn_ = {1595932860},
booktitle = {ACM MobiCom},
address_ = {New York, NY, USA},
url_ = {https://doi.org/10.1145/1161089.1161128},
doi_ = {10.1145/1161089.1161128},
abstract_ = {Although most wireless terrestrial networks are based on two-dimensional (2D) design, in reality, such networks operate in three-dimensions (3D). Since most often the size (i.e., the length and the width) of such terrestrial networks is significantly larger than the differences in the third dimension (i.e., the height) of the nodes, the 2D assumption is somewhat justified and usually it does not lead to major inaccuracies. However, in some environments, this is not the case; the underwater, atmospheric, or space communications being such apparent examples. In fact, recent interest in underwater acoustic ad hoc and sensor networks hints at the need to understand how to design networks in 3D. Unfortunately, the design of 3D networks is surprisingly more difficult than the design of 2D networks. For example, proofs of Kelvin's conjecture and Kepler's conjecture required centuries of research to achieve breakthroughs, whereas their 2D counterparts are trivial to solve. In this paper, we consider the coverage and connectivity issues of 3D networks, where the goal is to find a node placement strategy with 100\% sensing coverage of a 3D space, while minimizing the number of nodes required for surveillance. Our results indicate that the use of the Voronoi tessellation of 3D space to create truncated octahedral cells results in the best strategy. In this truncated octahedron placement strategy, the transmission range must be at least 1.7889 times the sensing range in order to maintain connectivity among nodes. If the transmission range is between 1.4142 and 1.7889 times the sensing range, then a hexagonal prism placement strategy or a rhombic dodecahedron placement strategy should be used. Although the required number of nodes in the hexagonal prism and the rhombic dodecahedron placement strategies is the same, this number is 43.25\% higher than the number of nodes required by the truncated octahedron placement strategy. We verify by simulation that our placement strategies indeed guarantee ubiquitous coverage. We believe that our approach and our results presented in this paper could be used for extending the processes of 2D network design to 3D networks.},
booktitle_ = {Proceedings of the 12th Annual International Conference on Mobile Computing and Networking},
pages_ = {346–357},
numpages_ = {12},
keywords_ = {3D networks, Kelvin's conjecture, connectivity, coverage, hexagonal prism, polyhedron, rhombic dodecahedron, three-dimensional networks, truncated octahedron, underwater networks, wireless networks},
location_ = {Los Angeles, CA, USA},
series_ = {MobiCom '06}
}

@misc{3gpp.23.003,
  author       = {{3GPP}},
  title_        = {{Numbering, Addressing and Identification}},
  institution_  = {{3rd Generation Partnership Project (3GPP)}},
  type_         = {Technical Specification (TS)},
  number_       = {23.003},
  version_      = {18.5.0},
  howpublished = {3GPP TS 23.003 v18.5.0, Release 18},
  year        = {2024},
  month_        = may,
  day_          = {21},
  note_         = {Version 18.5.0}
}

@misc{3gpp.29.118,
  author       = {{3GPP}},
  title        = {{Visitor Location Register (VLR); SGs Interface Specification}},
  institution_  = {{3rd Generation Partnership Project (3GPP)}},
  type_         = {Technical Specification},
  number_       = {29.118},
  howpublished = {3GPP TS 29.118 v18.0.0, Release 18},
  year         = {2024},
  month_        = may,
  day_          = {21},
  note_         = {Version 18.0.0, Release 18}
}

@misc{3gpp.36.321,
 author = {{3GPP}},
 institution_ = {{3rd Generation Partnership Project (3GPP)}},
 month_ = {05},
 note_ = {Version 18.1.0},
 number_ = {36.321},
 title = {{Medium Access Control (MAC) Protocol Specification}},
 type_ = {Technical Specification (TS)},
 howpublished = {3GPP TS 36.321 v18.1.0, Release 18},
 year = {2024}
}

@inproceedings{location_privacy_lte,
  title={{Location Privacy in LTE: A Case Study on Exploiting the Cellular Signaling Plane's Timing Advance}},
  author={John D. Roth and Murali Tummala and John C. McEachen and James W. Scrofani},
  booktitle={HICSS},
  year={2017},
  url_={https://api.semanticscholar.org/CorpusID:6152622}
}

@misc{ieee.802.11.2020,
 author = {{IEEE}},
 institution_ = {{IEEE Std 802.11™-2020, IEEE Standard for Information Technology}},
 title = {{Wireless LAN Medium Access Control (MAC) and Physical Layer (PHY) Specifications}},
 type_ = {Technical Specification (TS)},
 howpublished = {IEEE Std 802.11-2020},
 year = {2020}
}

@misc{ieee_wifi,
  author={{IEEE}},
  journal_={IEEE Std 802.11ax-2021}, 
  title={{Wireless LAN MAC and PHY Specifications, Amendment 1: Enhancements for High-Efficiency WLAN}}, 
  howpublished = {IEEE Std 802.11ax-2021},
  year={2021},
  volume_={},
  number_={},
  pages_={1-767},
  keywords_={IEEE Standards;Wireless LAN;Urban areas;Physical layer;Media Access Protocol;Information technology;Information exchange;Local area networks;Metropolitan area networks;dense deployment;high efficiency;IEEE 802.11™;IEEE 802.11ax™;MAC;medium access control;OFDMA;orthogonal frequency division multiple access;PHY;physical layer;wireless local area network;WLAN},
  doi_={10.1109/IEEESTD.2021.9442429}}

@inproceedings{monaco_sumo,
  year =        {2018},
  title =       {{M}onaco {SUMO} {T}raffic ({M}o{ST}) {S}cenario: {A} 3{D} {M}obility {S}cenario for {C}ooperative {ITS}},
  author =      {{C}odeca, {L}ara and  {H}{\"a}rri, {J}{\'e}r{\^o}me},
  booktitle =   {{SUMO} {U}ser {C}onference},
  address_ =     {{B}erlin, {Germany}},
  month_ =       {05},
  biburl_ =     {https://github.com/lcodeca/MoSTScenario}
}

@misc{amarisoft_callbox,
  author       = {Amarisoft},
  title        = {{Amarisoft Callbox Classics}},
  year         = {2024},
    month   = {11}
}

@inproceedings{SUMO2018,
          title = {{Microscopic Traffic Simulation using SUMO}},
         author = {Pablo Alvarez Lopez and Michael Behrisch and Laura Bieker-Walz and Jakob Erdmann and Yun-Pang Fl{\"o}tter{\"o}d and Robert Hilbrich and Leonhard L{\"u}cken and Johannes Rummel and Peter Wagner and Evamarie Wie{\ss}ner},
      publisher_ = {IEEE},
      booktitle = {IEEE ITSC},
           year = {2018},
        journal_ = {IEEE ITSC},
       keywords_ = {traffic simulation, modelling, optimization},
            url_ = {https://elib.dlr.de/124092/}
}

@article{wifi_loc_1,
  author       = {Pavel Kriz and
                  Filip Maly and
                  Tomas Kozel},
  title        = {{Improving Indoor Localization Using Bluetooth Low Energy Beacons}},
  journal      = {Mob. Inf. Syst.},
  volume_       = {2016},
  pages_        = {2083094:1--2083094:11},
  year         = {2016},
  url_= {https://doi.org/10.1155/2016/2083094},
  doi_= {10.1155/2016/2083094},
  timestamp_= {Mon, 26 Oct 2020 08:30:59 +0100},
  biburl_       = {https://dblp.org/rec/journals/mis/KrizMK16.bib},
  bibsource_= {dblp computer science bibliography, https://dblp.org}
}

@Article{ble_loc_1,
AUTHOR = {Cannizzaro, Davide and Zafiri, Marina and Jahier Pagliari, Daniele and Patti, Edoardo and Macii, Enrico and Poncino, Massimo and Acquaviva, Andrea},
TITLE = {{A Comparison Analysis of BLE-Based Algorithms for Localization in Industrial Environments}},
JOURNAL = {Electronics},
VOLUME_ = {9},
YEAR = {2020},
NUMBER_ = {1},
ARTICLE-NUMBER_ = {44},
url_ = {https://www.mdpi.com/2079-9292/9/1/44},
ISSN_ = {2079-9292},
}

@inproceedings{ble_loc_2,
author = {Park, Joonghong and Kim, Jaehoon and Kang, Sungwon},
year = {2015},
month = {12},
pages_ = {173-181},
title = {{BLE-Based Accurate Indoor Location Tracking for Home and Office}},
volume_ = {5},
booktitle = {CSIT},
doi_ = {10.5121/csit.2015.51614}
}

@inproceedings{ble_loc_3,
author = {Zafari, Faheem and Papapanagiotou,Ioannis},
year = {2015},
month_ = {12},
pages_ = {1-7},
title = {{Enhancing iBeacon Based Micro-Location with Particle Filtering}},
booktitle= {IEEE GLOBECOM},
doi_ = {10.1109/GLOCOM.2015.7417504}
}

@Article{ble_loc_4,
AUTHOR = {Wen, Xiaoyang and Tao, Wenyuan and Own, Chung-Ming and Pan, Zhenjiang},
TITLE = {{On the Dynamic RSS Feedbacks of Indoor Fingerprinting Databases for Localization Reliability Improvement}},
JOURNAL = {Sensors},
VOLUME_ = {16},
YEAR = {2016},
NUMBER_ = {8},
ARTICLE-NUMBER = {1278},
url_ = {https://www.mdpi.com/1424-8220/16/8/1278},
PubMed_ = {27537879},
ISSN = {1424-8220},
abstract_ = {Location data is one of the most widely used context data types in context-aware and ubiquitous computing applications. To support locating applications in indoor environments, numerous systems with different deployment costs and positioning accuracies have been developed over the past decade. One useful method, based on received signal strength (RSS), provides a set of signal transmission access points. However, compiling a remeasurement RSS database involves a high cost, which is impractical in dynamically changing environments, particularly in highly crowded areas. In this study, we propose a dynamic estimation resampling method for certain locations chosen from a set of remeasurement fingerprinting databases. Our proposed method adaptively applies different, newly updated and offline fingerprinting points according to the temporal and spatial strength of the location. To achieve accuracy within a simulated area, the proposed method requires approximately 3% of the feedback to attain a double correctness probability comparable to similar methods; in a real environment, our proposed method can obtain excellent 1 m accuracy errors in the positioning system.},
doi_ = {10.3390/s16081278}
}

@Article{ble_loc_5,
AUTHOR = {R{\"o}besaat, Jenny and Zhang, Peilin and Abdelaal, Mohamed and Theel, Oliver},
TITLE = {{An Improved BLE Indoor Localization with Kalman-Based Fusion: An Experimental Study}},
JOURNAL = {Sensors},
VOLUME_ = {17},
YEAR = {2017},
NUMBER_ = {5},
ARTICLE-NUMBER_ = {951},
url_ = {https://www.mdpi.com/1424-8220/17/5/951},
PubMedID_ = {28445421},
ISSN_ = {1424-8220},
abstract_ = {Indoor positioning has grasped great attention in recent years. A number of efforts have been exerted to achieve high positioning accuracy. However, there exists no technology that proves its efficacy in various situations. In this paper, we propose a novel positioning method based on fusing trilateration and dead reckoning. We employ Kalman filtering as a position fusion algorithm. Moreover, we adopt an Android device with Bluetooth Low Energy modules as the communication platform to avoid excessive energy consumption and to improve the stability of the received signal strength. To further improve the positioning accuracy, we take the environmental context information into account while generating the position fixes. Extensive experiments in a testbed are conducted to examine the performance of three approaches: trilateration, dead reckoning and the fusion method. Additionally, the influence of the knowledge of the environmental context is also examined. Finally, our proposed fusion method outperforms both trilateration and dead reckoning in terms of accuracy: experimental results show that the Kalman-based fusion, for our settings, achieves a positioning accuracy of less than one meter.},
doi_ = {10.3390/s17050951}
}

@Article{wifi_loc_3,
author={Du, Hongwei and Zhang, Chen and Ye, Qiang and Xu, Wen and Kibenge, Patricia Lilian
and Yao, Kang},
title={{A hybrid outdoor localization scheme with high-position accuracy and low-power consumption}},
journal={EURASIP J. Wirel. Commun. Netw.},
year={2018},
month_={Jan},
day_={04},
volume_={2018},
number_={1},
pages_={4},
issn_={1687-1499},
doi_={10.1186/s13638-017-1010-4},
url_={https://doi.org/10.1186/s13638-017-1010-4}
}

@inproceedings{FreudigerSH11,
  author       = {Julien Freudiger and
                  Reza Shokri and
                  Jean{-}Pierre Hubaux},
  editor_       = {George Danezis},
  title        = {{Evaluating the Privacy Risk of Location-Based Services}},
  booktitle_    = {Financial Cryptography and Data Security - 15th International Conference,
                  {FC} 2011, Gros Islet, St. Lucia, February 28 - March 4, 2011, Revised
                  Selected Papers},
  booktitle    = {FC},
  series__       = {Lecture Notes in Computer Science},
  volume_       = {7035},
  pages_        = {31--46},
  publisher_    = {Springer},
  year         = {2011},
  url_          = {https://doi.org/10.1007/978-3-642-27576-0\_3},
  doi_          = {10.1007/978-3-642-27576-0\_3},
  timestamp_= {Tue, 14 May 2019 10:00:38 +0200},
  bibsource_= {dblp computer science bibliography, https://dblp.org}
}

@inproceedings{Krumm07,
  author       = {John Krumm},
  editor_       = {Anthony LaMarca and
                  Marc Langheinrich and
                  Khai N. Truong},
  title        = {{Inference Attacks on Location Tracks}},
  booktitle    = {{PERVASIVE}},
  series__       = {Lecture Notes in Computer Science},
  volume_       = {4480},
  pages_        = {127--143},
  publisher_    = {Springer},
  year         = {2007},
  url_          = {https://doi.org/10.1007/978-3-540-72037-9\_8},
  doi_          = {10.1007/978-3-540-72037-9\_8},
  timestamp_= {Tue, 14 May 2019 10:00:36 +0200},
  bibsource_= {dblp computer science bibliography, https://dblp.org}
}

@misc{GAEN,
  author      = {Google and Apple},
  title       = {{Exposure Notification -- Bluetooth Specification}},
  year        = {2020},
  note        = {Version 1.2.2},
  url_         = {https://storage.googleapis.com/gweb-uniblog-publish-prod/documents/Exposure_Notification_-_Bluetooth_Specification_v1.2.2.pdf},
}

@inproceedings {sigover1,
author = {Hojoon Yang and Sangwook Bae and Mincheol Son and Hongil Kim and Song Min Kim and Yongdae Kim},
title = {{Hiding in Plain Signal: Physical Signal Overshadowing Attack on {LTE}}},
booktitle = {{USENIX Sec}},
year = {2019},
isbn_ = {978-1-939133-06-9},
address_ = {Santa Clara, CA},
pages_ = {55--72},
url_ = {https://www.usenix.org/conference/usenixsecurity19/presentation/yang-hojoon},
publisher_ = {USENIX Association},
month_ = {aug}
}

@inproceedings {Bae22_watchingthewatcher,
author = {Sangwook Bae and Mincheol Son and Dongkwan Kim and CheolJun Park and Jiho Lee and Sooel Son and Yongdae Kim},
title = {Watching the Watchers: Practical Video Identification Attack in {LTE} Networks},
booktitle = {{USENIX Sec}},
year = {2022},
isbn_ = {978-1-939133-31-1},
address_ = {Boston, MA},
pages_ = {1307--1324},
url_= {https://www.usenix.org/conference/usenixsecurity22/presentation/bae},
publisher_ = {USENIX Association},
month_ = aug
}

@misc{3gpp.36.331.18.5.0,
  author       = {{3GPP}},
  title        = {{LTE; Evolved Universal Terrestrial Radio Access (E-UTRA); Radio Resource Control (RRC); Protocol Specification}},
  institution_  = {{3rd Generation Partnership Project (3GPP)}},
  type_         = {Technical Specification},
  number_       = {36.331},
  howpublished = {3GPP TS 36.331 v18.5.0, Release 18},
  year         = {2025},
  month_        = mar,
  day_          = {21},
  note_         = {Version 18.0.5, Release 18}
}

@inproceedings{ndss_deanonymization_ble_wifi,
  author = {Christopher Ellis and Yue Zhang and Mohit Kumar Jangid and Shixuan Zhao and Zhiqiang Lin},
  title  = {Deanonymizing Device Identities via Side-channel Attacks in Exclusive-use IoTs {\&} Mitigation},
  booktitle    = {{NDSS}},
  publisher_    = {The Internet Society},
  year         = {2025},
  url_          = {https://www.ndss-symposium.org/ndss-paper/deanonymizing-device-identities-via-side-channel-attacks-in-exclusive-use-iots-mitigation/},
  timestamp_    = {Wed, 19 Mar 2025 15:12:36 +0100},
  biburl_       = {https://dblp.org/rec/conf/ndss/Ellis0J0025.bib},
  bibsource_    = {dblp computer science bibliography, https://dblp.org}
}

@inproceedings{inria_ble,
  author       = {Lo{\"{\i}}c Jouans and
                  Aline Carneiro Viana and
                  Nadjib Achir and
                  Anne Fladenmuller},
  title        = {Associating the Randomized Bluetooth {MAC} Addresses of a Device},
  booktitle    = {IEEE CCNC},
  pages_        = {1--6},
  publisher_    = {{IEEE}},
  year         = {2021},
  url_          = {https://doi.org/10.1109/CCNC49032.2021.9369628},
  doi_          = {10.1109/CCNC49032.2021.9369628},
  timestamp_    = {Mon, 03 Mar 2025 20:59:24 +0100},
  biburl_       = {https://dblp.org/rec/conf/ccnc/JouansVAF21.bib},
  bibsource_    = {dblp computer science bibliography, https://dblp.org}
}

@article{Bao2022,
  author = {Fanchen Bao and Stepan Mazokha and Jason O. Hallstrom},
  title = {Mobility Intelligence: Machine Learning Methods for Received Signal Strength Indicator-based Passive Outdoor Localization},
  journal = {Adv. Sci. Technol. Eng. Syst. J.},
  year = {2022},
  volume_ = {7},
  number_ = {6},
  pages_ = {269–282},
  doi_ = {10.25046/aj070631},
  url_ = {https://www.astesj.com/v07/i06/p31/},
  language_ = {en},
  publisher_ = {ASTES Publishers},
  
}

@inproceedings{etzlinger_ble_distance,
  author    = {Bernhard Etzlinger and
               Barbara Nu{\ss}baumm{\"u}ller and
               Philipp Peterseil and
               Karin Anna Hummel},
  title     = {{Distance Estimation for {BLE}-Based Contact Tracing: A Measurement Study}},
  booktitle = {IEEE Wireless Days},
  pages_     = {1--5},
  year      = {2021},
  publisher_ = {IEEE},
  doi_       = {10.1109/WD52248.2021.9508280}
}

@article{Bao2025Streetscape,
  author  = {Fanchen Bao and Stepan Mazokha and Jason O. Hallstrom},
  title   = {{RSSI}-Based Passive Localization in the Wild, at Streetscape Scales},
  journal = {IEEE J. Indoor Seamless Position. Navig.},
  year    = {2025},
  doi_     = {10.1109/JISPIN.2025.3534200}
}

\appendix
\section{Ethical Considerations}
Over the course of doing the research we were very careful about how we collected data. The real-world experiments we did were all executed in carefully controlled lab environments. We did not eavesdrop on radio signals of arbitrary device of other people, nor did we measure or try to recover movement patterns of real people. To evaluate the effectiveness of our approaches, we instead used our lab measurements combined with real-world logs of our own devices to understand protocol behavior.

\rev{Our results show that current per-protocol randomization is not sufficient to prevent tracking by adversaries. \name does not exploit any specific patchable flaw in any single implementation. It is a consequence of independently standardized protocols: LTE, WiFi, and BLE. Even when each protocol behaves exactly as specified, the privacy loss arises from their unsynchronized randomization. Adversaries with multi-protocol sniffing capabilities could use our work to track users more easily. However, protocol-specific flaws already make users easy to track. We argue that encouraging research into better cross-protocol and cross-user countermeasures outweighs this potential harm. Because the attack involves multiple protocols, there is no single entity to whom we can disclose this vulnerability. We will nonetheless report our findings to 3GPP, IEEE, and the Bluetooth SIG. We have reported the vulnerability to both major mobile OS vendors, Google and Apple, are awaiting their response. We have recommended them to synchronize identifiers within the same device, similar to DP3T/GAEN, such that linking between the identifier of the same device could be evaded.}

\section{Open Science}
We commit to following ACM CCS's open science policy. The artifact is available at: \url{https://doi.org/10.5281/zenodo.22499946}

Our artifacts consist of:
\begin{enumerate}
    \item A dataset of real commodity device capture and scripts to run the CrossLink to generate Figure~\ref{fig:real_devices}.
    \item A code artifact that implements our prototype tracking algorithm from Section~\ref{sec:algorithm}.
    \item A code artifact that simulates movements of real-world users based on two sources: (a) a simulation based on SuMo~\cite{SUMO2018} movement patterns in Monaco~\cite{monaco_sumo}; (b) synthetic simulations on a grid.
    \item A dataset of simulated user movement patterns: for the different Monaco and synthetic settings we use in the paper. (These were generated from (3), we include the real data to avoid issues with randomness.)
    \item A dataset of a collection of simulated user device transmissions. We include a set for each movement pattern (4) and set of protocol parameters (as determined in (1)) to cover all configurations evaluated in Section~\ref{sec:evaluation} (e.g., Figure~\ref{fig:privacy_leakage_q1}). We include these datasets to avoid issues with randomness.
    \item A code artifact that computes the observations by a passive eavesdropper given the transmissions of users. This artifact has configurations to reproduce sniffer placements (Figure~\ref{fig:privacy_leakage_q2}).
    \item The evaluation scripts that take simulated eavesdropping observations (these are uniquely determined by dataset (4) and artifact+configurations (5)) and the prototype tracking scripts (2) that reproduce and evaluate the tracking results presented in the figures in Section~\ref{sec:evaluation}.
    
\end{enumerate}

\fullonly{\section{Generative AI Usage}}

\fullonly{The authors used Claude and Grammarly to assist with editing and paraphrasing selected portions of the manuscript for clarity and presentation. In particular, AI assistance was used for text refinement in parts of Section~\ref{section:characterization} to make it more concise. The results were carefully checked by the authors. Grammarly was used for refinement throughout the paper for minor editorial changes. No AI tools were used to generate experimental data, implement the evaluation, compute results, or formulate the paper's technical claims.}

\fullonly{
\section{Mathematical Model of Mixing}
\label{app:bounds}

\subsection{Analysis of Theoretical Results}
In a typical setting (see also Section~\ref{sec:simulation_setup}) we would expect the randomization rate $\lambda_r$ to be much smaller than the transmission rate $\lambda_t$. An example setting for BLE could be $\lambda_r = 1/600 s^{-1}$ and $\lambda_t = 1$. 

We see several expected effects occur. The smaller $T$ (i.e., the time that the second device has to also transmit its randomized identifier) the smaller the upper bound, and thus the less likely the devices are to mix. Similarly, increasing $\lambda_t$, the transmission rate, results in smaller upper bound. Finally, decreasing the rate of randomization $\lambda_r$ also results in a smaller upper bound.

For practical values of $\lambda_r$ and $\lambda_t$ this upper bound is concretely small. For example, with the values mentioned just now, $\probmix \leq 0.0034$. As the time $t$ increases in which the two devices are close, the more often a device randomizes its identifier, and the more opportunities for mixing exist. However, opportunities are rare, and the probability of mixing successfully at such an opportunity is even rarer.

This problem is exacerbated when considering multiple protocols. In this setting, when two devices are close, for successful randomization to happen, \emph{all} protocol identifiers must mix. This makes an already rare event even rarer.

\subsection{Proof of Theorem~\ref{thm:mix-lower_bound_multi}}
We define the following conditions necessary to achieve perfect mixing for a given protocol. We assume that the randomization and transmission procedure of the protocols are independent from each other.

\begin{enumerate}
    \item \textbf{Identifier Randomization}: Both devices must randomize their identifiers within the mix-zone duration $T$, during which their location traces are indistinguishable. In this work, the mix-zone is defined as the time period when the location traces of two devices cannot be differentiated.
    \item \textbf{Identifier Transmission Order}: When one device randomizes its identifier, perfect mixing is achieved under one of the following conditions:
    \begin{enumerate}
        \item The first device transmits its randomized identifier after the second device randomizes its identifier.
        \item The first device transmits its randomized identifier before the second device randomizes its identifier, provided the second device does not transmit its original identifier during this interval.
        \end{enumerate}
\end{enumerate}

To measure the probability of perfect mixing, we model the randomization interval and transmission interval as exponential distributions with specified rate parameters. The exponential distribution is chosen because:

\begin{itemize}
    \item Transmission intervals in wireless communication are non-periodic and influenced by factors such as smartphone usage, notifications, and installed applications. Since the timing of the next transmission is independent of previous transmissions, the exponential distribution is an appropriate model.
    \item Randomization intervals depend on changes in access points, which are influenced by user mobility and are independent of past randomization events, making the exponential distribution suitable for this scenario.
\end{itemize}

To compute the probability of perfect mixing in a protocol, we define the random variables as follows:

\begin{itemize}
    \item $R_{1} \sim \text{Exponential}(\lambda_{r}):$  time instance of the randomization in device 1.
    \item $R_{2} \sim \text{Exponential}(\lambda_{r}):$ time instance of the randomization in device 2.
\end{itemize}

Now, focusing on the randomization condition, we compute the probability of both the devices randomizing their identifier within the mix-zone duration $T$.

 \begin{align} 
{\sf Pr}(R_{1}\leq T \land R_{2} \leq T)=
(1-\exp(-\lambda_{r}T))^2
\end{align}

Since both $R_1$ and $R_2$ are drawn from the same distribution, the probability that $R_{1} < R_{2}$ is $1/2$. Since, for perfect mixing, the transmission restrictions depend on the order of randomization, we define the random variables corresponding to the transmission events as follows, conditioned on $R_{1}<R_{2}$.

\begin{itemize}
\item $T_{1}$: time instance at which device $1$ identifier is first transmitted after randomization: $T_{1}=R_{1}+X_{1}$, where $X_1 \sim \text{Exponential}(\lambda_{t})$
\item  $T_{2}$: time instance at which device 2 identifier is first transmitted after the first transmission of device 1 
identifier after randomization: $T_{2}=R_{1}+X_{1}+X_{2}$, where, $X_{2} \sim \text{Exponential}(\lambda_{t})$
\end{itemize}

\begin{table}[hbtp]

\centering
\begin{tabular}{|l|l|l|}
\hline
$R_{1}<R_{2}$    & $T_{2}<R_{2}$ & $T_{2}>R_{2}$ \\ \hline
$T_{1}<R_{2}$    & $\times$       & $\checkmark$      \\ \hline
$T_{1}>R_{2}$ & $\checkmark$   & $\checkmark$      \\ \hline
\end{tabular}
\caption{Logical conditions for the relationship between transmission ($T_1$,$T_2$) and randomization ($R_1,R_2$) events. The table highlights the conditions ($\checkmark$ for valid and $\times$ for invalid) for perfect mixing based on comparisons between the randomization and transmission events conditioned on $R_{1}<R_{2}$.}
\label{tab:p_mix}
\end{table}

We use these random variables to analyze the cases where mixing does and does not happen, see Table~\ref{tab:p_mix}. By definition the case $T_1 > R_2$ and $T_2 < R_2$ cannot happen (because $T_2 \geq T_1$). Hence,
\begin{multline}
    {\sf Pr}(\text{mix} \;|\; R_{1} < R_{2}) = {\sf Pr}(R_{1}\leq T \land R_{2} \leq T) \cdot \\ {\sf Pr}(T_{2}>R_{2}|R_{1}\leq T \land R_{2}\leq T)
\end{multline} 
We compute both terms.
Since, the randomization of the devices are independent from one another, 
\begin{align*}
    {\sf Pr}(R_{1}\leq T \land R_{2} \leq T) 
    &= (1-e^{-\lambda_{r}T})(1-e^{-\lambda_{r}T}) \\ &= (1-e^{-\lambda_{r}T})^2
\end{align*}

Now, we compute the ${\sf Pr}(T_{2}>R_{2}|R_{2}\leq T \land R_{1}\leq T)$. This requires tedious integrals, to eventually yield that:
\begin{multline*}
    {\sf Pr}(T_{2}>R_{2}|R_{2}\leq T \land R_{1}\leq T)
    =\frac{\lambda_{r}^2}{2\lambda_{r}}\{\frac{1}{B}+\frac{\lambda_{t}}{B^2}\}(1-e^{-2\lambda_{r}T}) \\
    -\lambda_{r}^2e^{-BT}\{(\frac{1}{B}  +\frac{\lambda_{t}}{B^2})(\frac{1-e^{-AT}}{A})+\frac{\lambda_{t}}{B}(\frac{T}{A}
    -\frac{1-e^{-AT}}{A^2})\}
\end{multline*}
Therefore the final probability of mixing is:
\begin{multline*}
    {\sf Pr}(\text{mix})
    =\lambda_{r}\{\frac{1}{B}+\frac{\lambda_{t}}{B^2}\}(1-e^{-2\lambda_{r}T})
    \\
    -2\lambda_{r}^2e^{-BT}\left[(\frac{1}{B}+\frac{\lambda_{t}}{B^2})(\frac{1-e^{-AT}}{A})
    +\frac{\lambda_{t}}{B}(\frac{T}{A}-\frac{1-e^{-AT}}{A^2})\right]\\
    \text{where }A=\lambda_{r}-\lambda_{t}\text{ and }B=\lambda_{r}+\lambda_{t}
\end{multline*}

In the above expression, we can see that the second term is multiplied with a factor of $\lambda_{r}^2 \cdot e^{-BT}$. Since, $\lambda_{r}<1$ and the term $e^{-BT}$ decreases exponentially with increasing $T$, the entire second term becomes negligible. Therefore, the final expression for the probability of perfect mixing at a single opportunity satisfies

\begin{align*}
    {\sf Pr}(\text{mix}) \leq \lambda_{r}\{\frac{1}{B}+\frac{\lambda_{t}}{B^2}\}(1-e^{-2\lambda_{r}T})
\end{align*}

In the above term, with $T \to \infty$, $(1-e^{-2\lambda_{r}T}) = 1$. Therefore, the probability of mixing in a particular protocol is strictly upper bounded by the following equation
\begin{align*}
    {\sf Pr}(\text{mix}) \leq \lambda_{r}\{\frac{1}{B}+\frac{\lambda_{t}}{B^2}\}. \qedhere
\end{align*}

Considering the upper bound of mixing after the first randomization, we denote it as $p_{ub}$. 

\begin{align*}
    {\sf Pr}(\text{no mix})\geq 1-p_{ub}
\end{align*}

Consider two devices, each performing randomization events independently, according to a Poisson process with the same rate parameter $\lambda_{r}$. Let $Y_{1}$ and $Y_{2}$ be the random variables counting the number of randomization events for each device. Thus, $Y_{1} \sim \text{Poisson}(\lambda_{r})$ and $Y_{2} \sim \text{Poisson}(\lambda_{r})$. Since each randomization event is a mixing opportunity, considering both the devices, the number of mixing opportunities will be denoted by $Y=Y_{1}+Y_{2}$, resulting in a Poisson distribution with rate parameter $Y \sim \text{Poisson}(2\lambda_{r})$. Therefore, within the time duration $T$, the expected number of mixing opportunities is $2\lambda_{r}T$.

We now find the upper bound on the probability of no mixing in none of the opportunities in the duration $T$.

\begin{align*}
      {\sf Pr}(\text{no mix in }T) & \geq \sum_{k=0}^{\infty}p(Y=k){\sf Pr}(\text{no mix}|Y=k)\\
     & \geq p(Y=k)(1-p_{ub})^{k}\\
     & \geq \frac{(2\lambda_{r}T)^{k}e^{-2\lambda_{r}T}}{k!}(1-p_{ub})^{k}\\
     & \geq e^{-2\lambda_{r}Tp_{ub}}
\end{align*}

Therefore, the upper bound on the probability of mixing in $T$ is determined by:

\begin{align}
    {\sf Pr}(\text{mix in }T) \leq 1-e^{-p_{ub}2\lambda_{r}T}
\end{align}
\FloatBarrier 

 \section{Derivation of the Error Expressions}
 \label{app:sigma-derivation}

 This section provides the derivations for the error expressions in heavy-tailed error model from Section~\ref{sec:simulation_setup}. For a true device to sniffer distance $d>0$, the estimated distance is
 \begin{align*}
      \widehat{d} &= d\eta, &
      \eta &= e^{\sigma Z}, &
      Z &\sim \mathcal{N}(0,1),
  \end{align*}

  where $\sigma>0$ controls the error magnitude. The signed distance error is therefore
  \begin{align*}
      E = \widehat{d}-d = d\eta-d = d(\eta-1).
  \end{align*}

 Let $\Phi$ denote the standard-normal cumulative distribution function (CDF). We derive the standard deviation, mean absolute error, and absolute-error quantile of $E$. All three statistics follow from the standard-normal moment identities: 
 \begin{equation}
     \mathbb{E}[e^{tZ}]=e^{t^{2}/2}.
      \label{eqn:normal-moment}
 \end{equation}
 and
 \begin{align}
 \mathbb{E}\!\left[
     e^{tZ}\mathbf{1}_{\{Z\geq0\}}
      \right]
      &=
      e^{t^{2}/2}\Phi(t),
     \label{eqn:normal-positive-moment}\\
      \mathbb{E}\!\left[
          e^{tZ}\mathbf{1}_{\{Z<0\}}
      \right]
      &=
      e^{t^{2}/2}\Phi(-t).
      \label{eqn:normal-negative-moment}
  \end{align}

 \noindent\textbf{Standard deviation.}
 Evaluating the moment-generating function at $t=\sigma$ and $t=2\sigma$ gives
 $\mathbb{E}[\eta]=e^{\sigma^{2}/2}$ and $\mathbb{E}[\eta^{2}]=e^{2\sigma^{2}}$.
 Since $E=d(\eta-1)$ only shifts by a constant and scales by $d$,
 \begin{equation}
     \operatorname{SD}(E)=d\sqrt{e^{\sigma^{2}}\!\left(e^{\sigma^{2}}-1\right)}.
     \label{eqn:error-sd}
 \end{equation}

 \noindent\textbf{Mean absolute error.}
 Because $\eta\geq1\iff Z\geq0$, the absolute value splits at $z=0$ and the two
 $\tfrac{1}{2}$ probability terms cancel by symmetry; the half-line identity
 (at $t=\sigma$ and its reflection) then gives
 $\mathbb{E}[|\eta-1|]=e^{\sigma^{2}/2}\!\left(\Phi(\sigma)-\Phi(-\sigma)\right)$,
 hence
 \begin{equation}
     \operatorname{MAE}(E)=d\,e^{\sigma^{2}/2}\left(2\Phi(\sigma)-1\right).
     \label{eqn:error-mae}
 \end{equation}

 \noindent\textbf{Absolute-error quantile.}
 For a threshold $0\leq a<d$, monotonicity of $\eta=e^{\sigma Z}$ gives
 $|E|\leq a\iff \ln(1-a/d)/\sigma\leq Z\leq \ln(1+a/d)/\sigma$, so
 \begin{equation}
     \Pr(|E|\leq a)
     =\Phi\!\left(\frac{\ln(1+a/d)}{\sigma}\right)
     -\Phi\!\left(\frac{\ln(1-a/d)}{\sigma}\right).
     \label{eqn:error-quantile}
 \end{equation}

For BLE, Etzlinger et al.~\cite{etzlinger_ble_distance} fit a log-distance path-loss model with path-loss exponent $\gamma=3.36$ and Gaussian path-loss noise with standard deviation $\sigma_{\mathrm{dB}}=8.48$\,dB. Mapping these fitted parameters to our multiplicative distance-error model gives $\sigma_{\mathrm{BLE}}
=(\ln 10)\sigma_{\mathrm{dB}}/(10\gamma)=0.58$. For WiFi, we average the four reported MAEs to obtain $4.04$\,m and match this value at a $15$\,m reference link, equal to the testbed width~\cite{Bao2022}. Solving the MAE
 expression gives $\sigma_{\mathrm{WiFi}}=0.326$. Finally, LTrack~\cite{ltrack} reports an approximately $6$\,m 90th-percentile error for commercial phones over distances up to $60$\,m~\cite{ltrack}. Setting $a_{0.9}=6$\,m, $d_{\mathrm{LTE}}^\star=60$\,m, and $q=0.9$ in the quantile expression gives $\sigma_{\mathrm{LTE}}=0.061$.

 \FloatBarrier

\section{Glossary}
\label{app:glossary}

To improve readability and enhance readers' understanding, we describe the notations in Table~\ref{tab:notation} and the acronyms in Table~\ref{tab:acronyms}.

\begin{table}[H]
\caption{Notation used in the algorithm and analysis.}
\label{tab:notation}
\small
\setlength{\tabcolsep}{5pt}
\begin{tabular}{@{}ll@{}}
\toprule
\textbf{Symbol} & \textbf{Meaning} \\
\midrule
$\mathcal{D}$ & A target device \\
$\mathcal{A}$ & the passive adversary \\
$\textsf{LP}_{\mathcal{D}}$ & Location Privacy Leakage of Device $\mathcal{D}$ (Definition~\ref{def:privacy}) \\
$\protid{}$ & A temporary identifier \\
$t$ & timestamp of an observation \\
$d$ & true distance between sniffer and device \\
$\dhat$ & approximated noisy distance between the sniffer and the device \\
$\mathcal{S}$ & set of sniffers \\
$\sniffer{i}$,$\snifloc{i}$ & the $i$-th sniffer and its corresponding location\\
$\epsilon_{p}$ & Protocol-specific error tolerance \\
$V_{max}$ & Assumed upper bound on device speed \\
$\maxiat{p}$ & Upper bound on inter-arrival message time for protocol $p$.\\
$\widehat{P}$ & set of monitored protocols \\
$\eta_p$ & Multiplicative log-normal ranging-error factor \\
$\sigma_p$ & Heavy-tail error-scale parameter \\
$Z$ & Standard-normal random variable\\
\bottomrule
\end{tabular}
\end{table}

\begin{table}[H]
\caption{Acronyms used throughout the paper.}
\label{tab:acronyms}
\small
\setlength{\tabcolsep}{6pt}
\renewcommand{\arraystretch}{1.05}
\begin{tabular}{@{}lp{0.70\columnwidth}@{}}
\toprule
\textbf{Acronym} & \textbf{Meaning} \\
\midrule
BDADDR  & Bluetooth Device Address \\
BLE     & Bluetooth Low Energy \\
BSR     & Buffer Status Report \\
BTC     & Bluetooth Classic \\
C-RNTI  & Cell Radio Network Temporary Identifier \\
IMSI    & International Mobile Subscriber Identity \\
MAC     & Medium Access Control (address) \\
TMSI    & Temporary Mobile Subscriber Identity \\
eNodeB  & Evolved Node B (LTE base station) \\
PCI     & Physical Cell Identity \\
PHR     & Power Headroom Report \\
QoS     & Quality of Service \\
RRC     & Radio Resource Control \\
SRB     & Signaling Radio Bearer \\
TA      & Timing Advance \\
UE & User Equipment \\
\bottomrule

\end{tabular}
\end{table}
\FloatBarrier

\section{Real Device Packet Captures}

This appendix section provides the packet captures behind the observations in
Section~\ref{section:characterization}: identifier linking during connection setup and
re-establishment (Figures~\ref{fig:connectionsetup}--\ref{fig:test_rrc_connection_request}),
identifier stability across reconfiguration and intra-eNodeB handover
(Figures~\ref{fig:rrc_connection_reconfig_no_change} and~\ref{fig:rrc_intra_enb}),
and our testbed (Figure~\ref{fig:setup}).

\begin{figure}[H]
    \centering
    \includegraphics[width=1\linewidth]{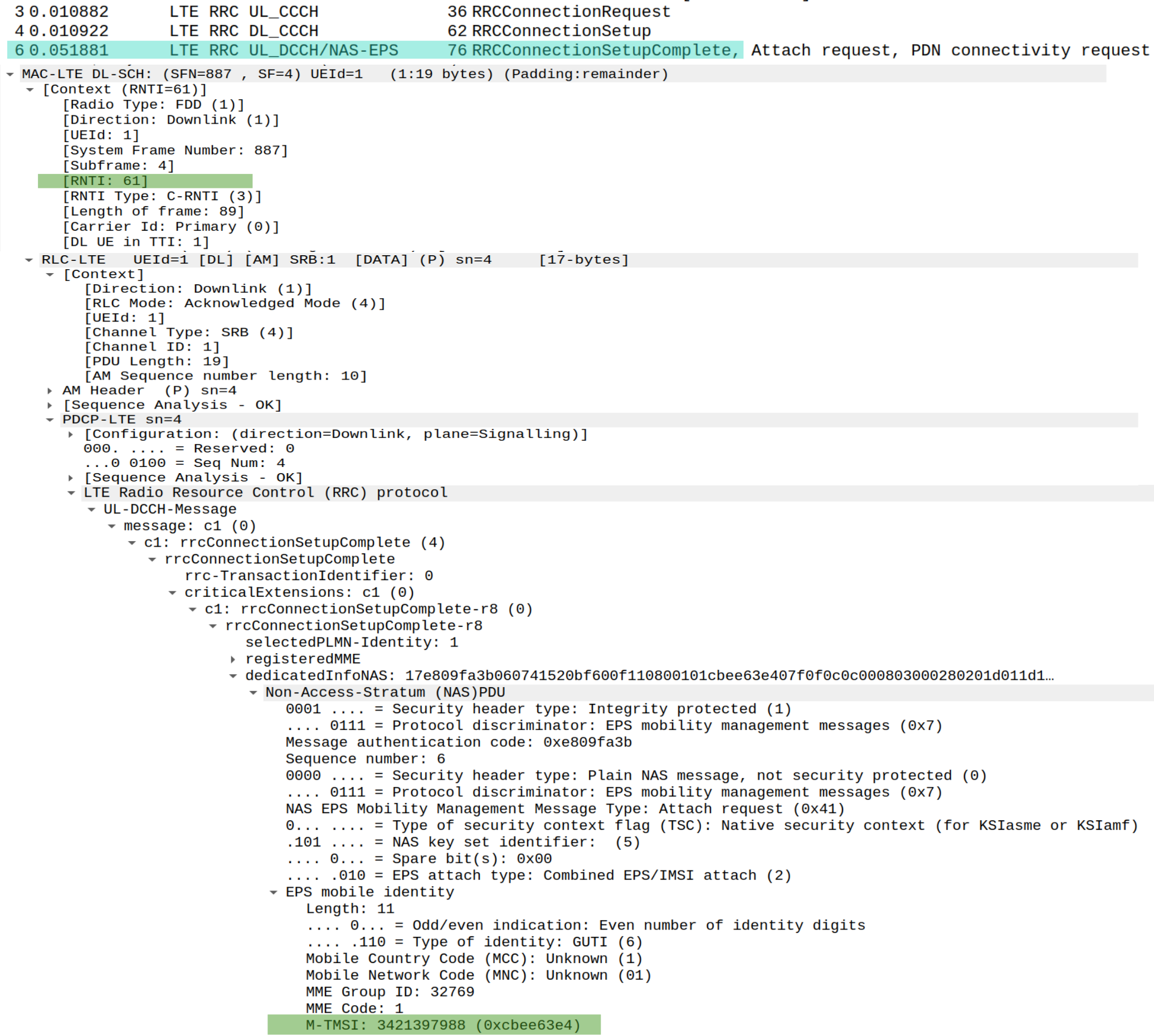}
    \caption{Linking between TMSI and C-RNTI value in RRC Connection Setup Phase}
    \label{fig:connectionsetup}
    \Description{RRC connections setup}
\end{figure}

\begin{figure}[H]
    \centering
    \includegraphics[width=1\linewidth]{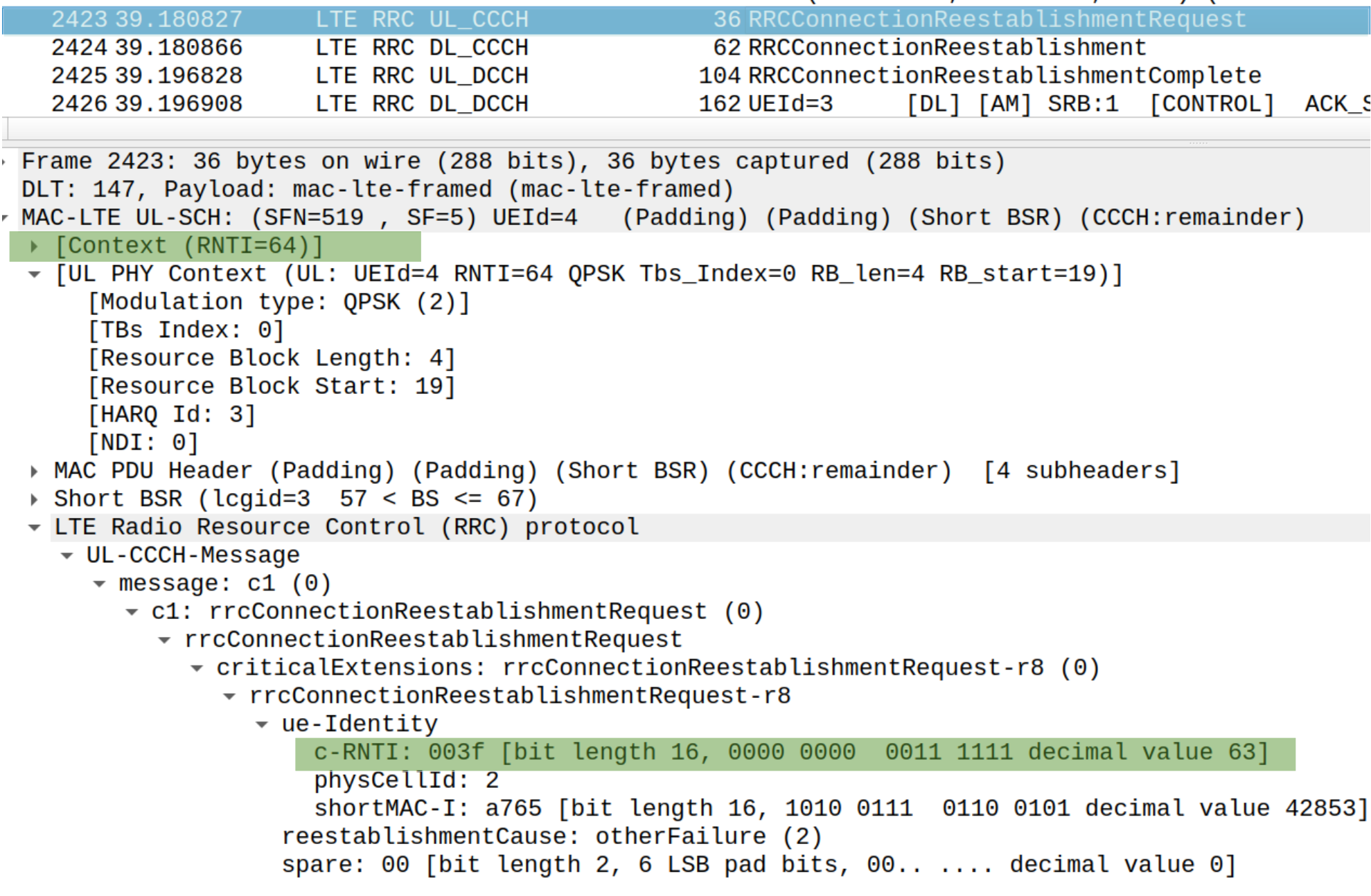}
    \caption{Linking between old C-RNTI value and new C-RNTI value during link failure.}
    \Description{Diagram showing how the old C-RNTI is linked to the new C-RNTI during link failure recovery.}
    \label{fig:rrc_reestablishment}
\end{figure}

\begin{figure}[H]
     \centering
     \includegraphics[width=1\linewidth]{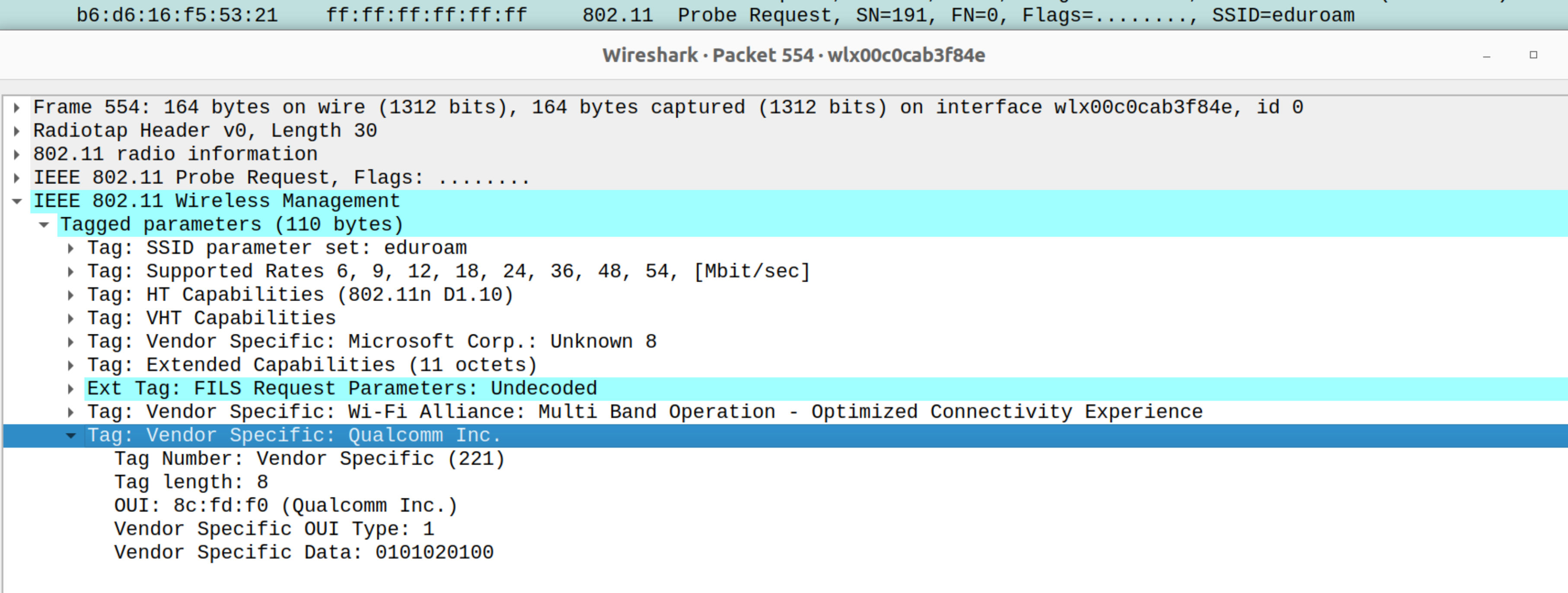}
     \caption{Tracking WiFi device with labels in Probe Requests like SSID, Vendor Specific tags}
     \Description{Probe request in Wifi}
     \label{fig:wifi_vendor}
 \end{figure}

\begin{figure}[H]
    \centering
    \includegraphics[width=1\linewidth]{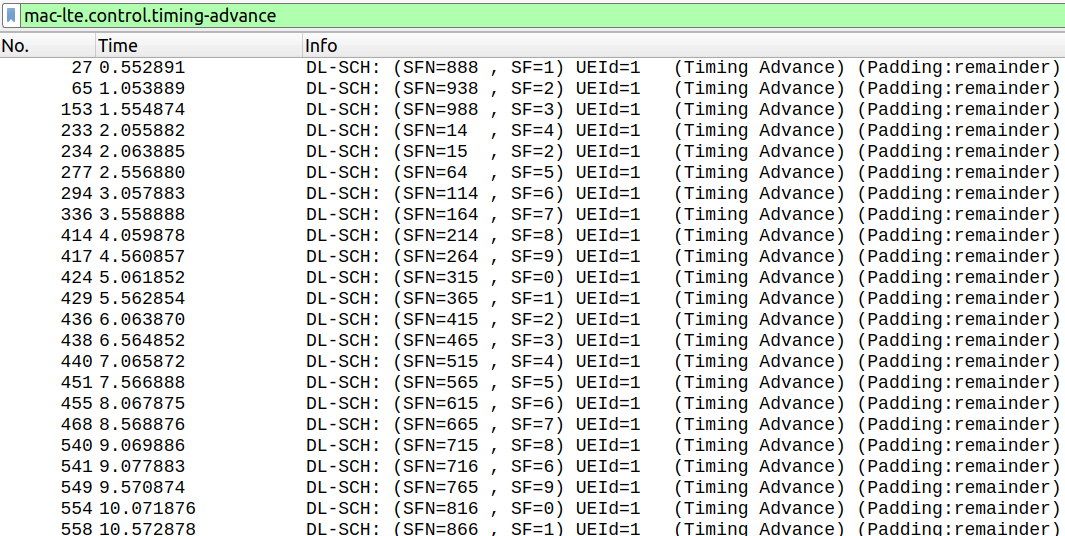}
    \caption{Timing Advance Obtained every 0.5 seconds interval in the window of 10 seconds}
    \Description{Frequency of timing advanced command}
    \label{fig:timing_advance}
\end{figure}

\begin{figure}[H]
    \centering
    \includegraphics[width=1\linewidth]{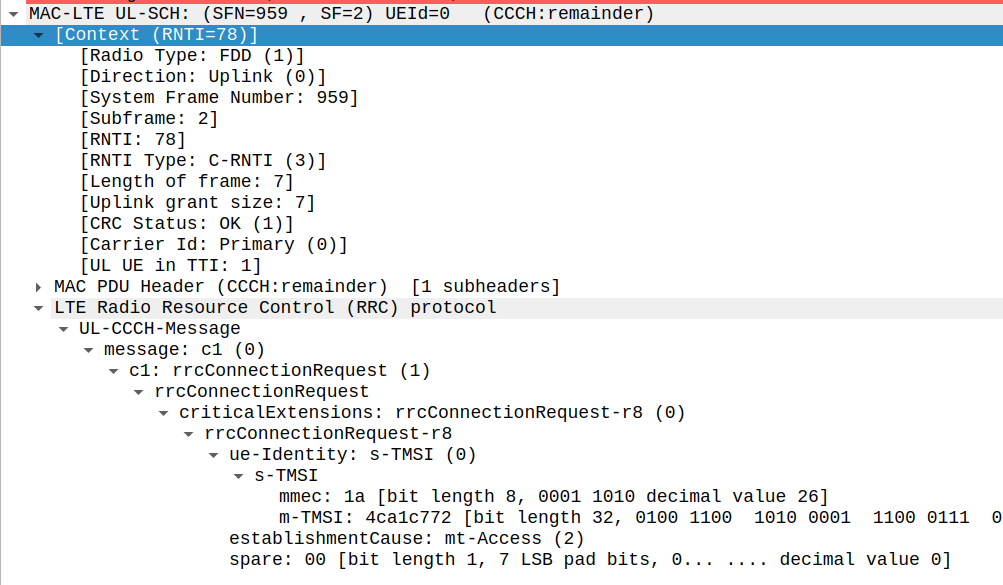}
    \caption{RRC Connection Request containing TMSI value along with C-RNTI during the connection re-initiation}
    \Description{C-RNTI-TMSI mapping in RRC connection request}
    \label{fig:test_rrc_connection_request}
\end{figure}

\begin{figure}[H]
    \centering
        \includegraphics[width=1\linewidth]{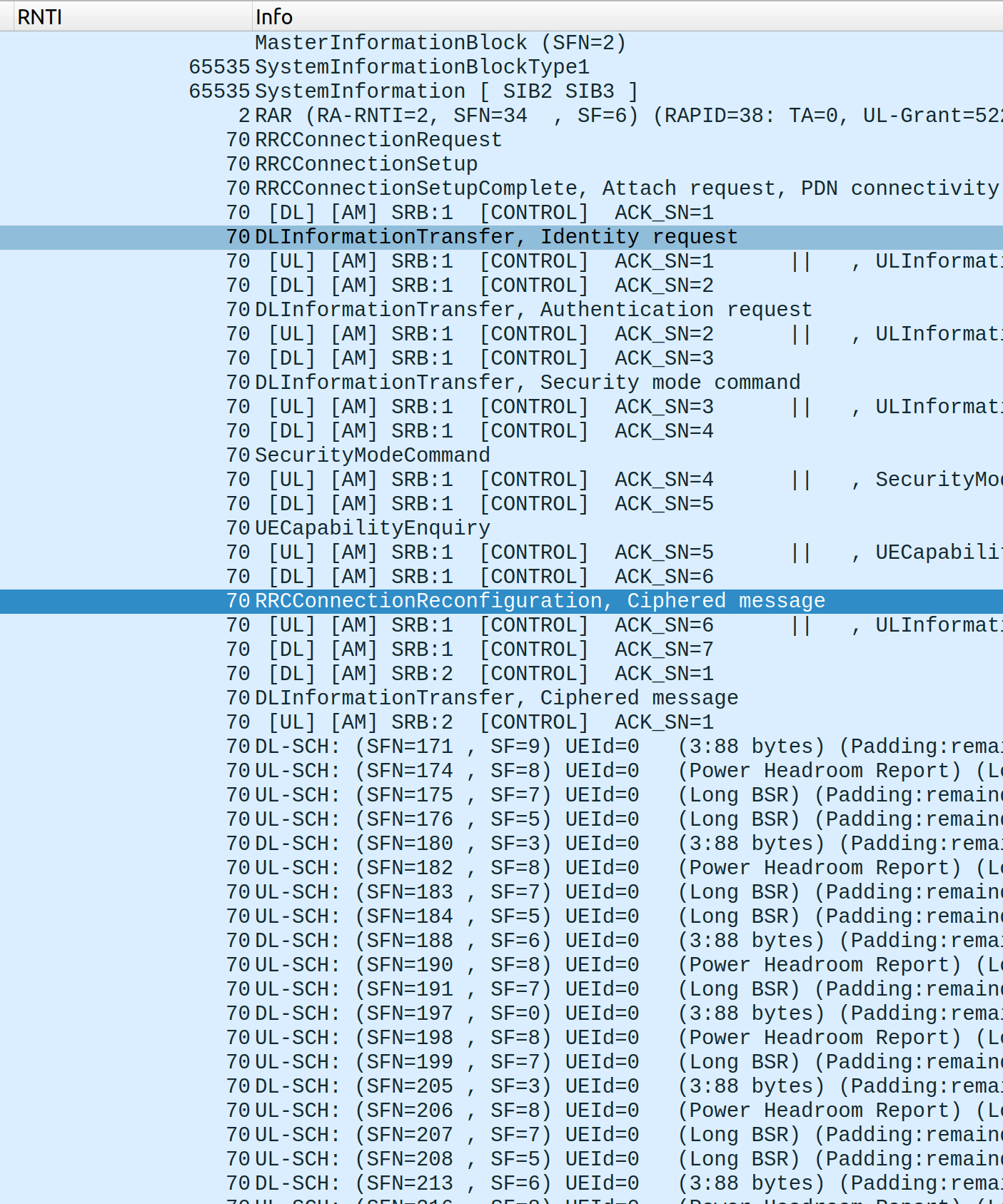}
    \caption{No change in C-RNTI value after reception of RRC Reconfiguration message when connected to the same base station}
    \Description{No change in C-RNTI when connected to same enodeB}
    \label{fig:rrc_connection_reconfig_no_change}
\end{figure}

\begin{figure}[H]
    \centering
        \includegraphics[width=1\linewidth]{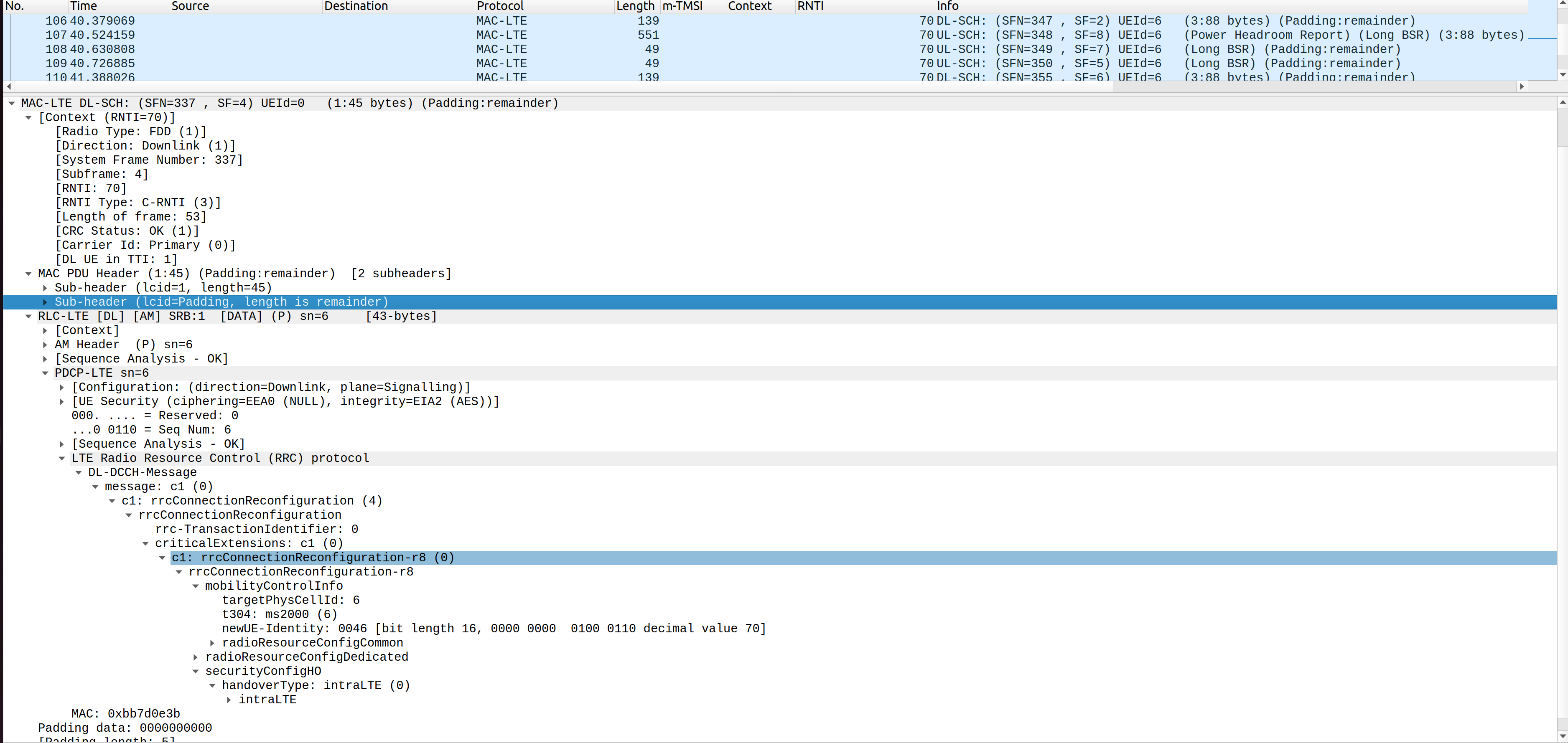}
    \caption{No change in C-RNTI value during intra-enodeB handover}
    \label{fig:rrc_intra_enb}
    \Description{intra enodeB handover}
\end{figure}

\begin{figure} [H]
    \centering
    \includegraphics[width=0.9\linewidth]{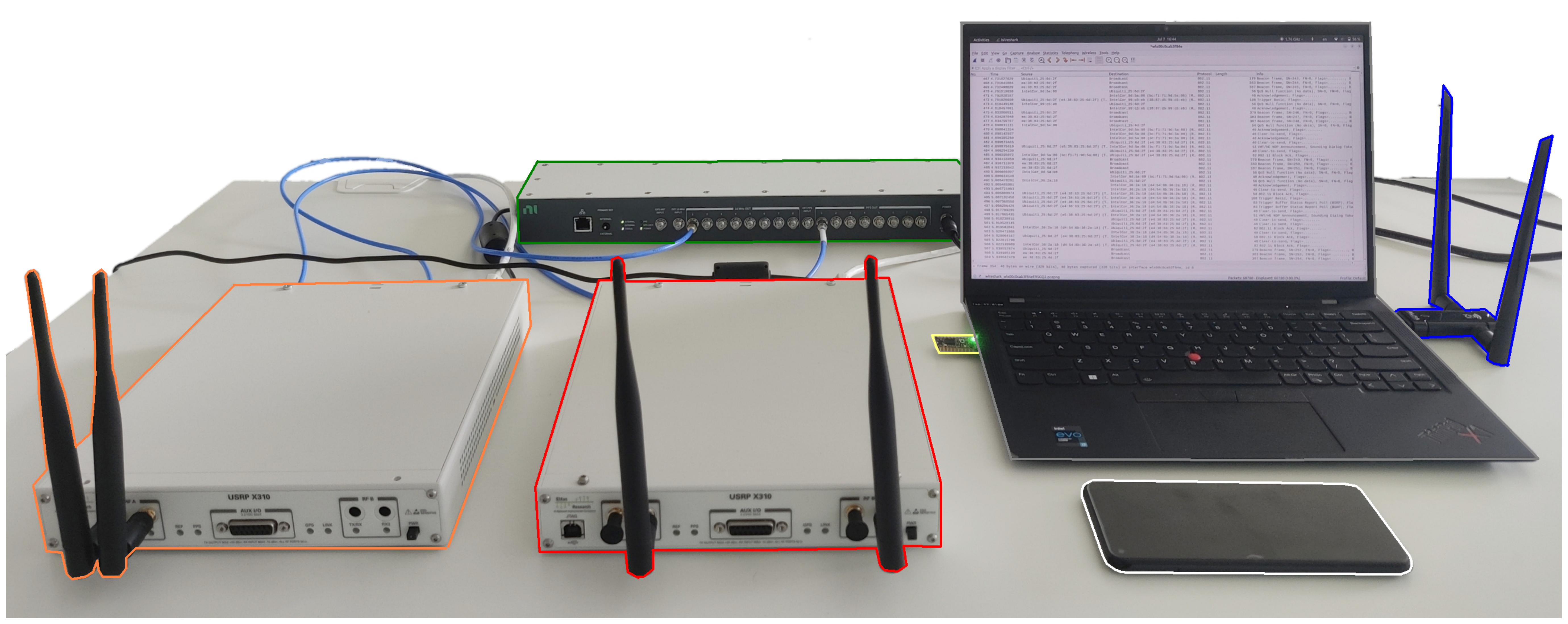}
    \caption{Our experimental setup. Devices are color coded. From left to right: an USRP X310 as eNodeB (orange); another USRP X310 as a sniffer (red); a laptop (connected to the USRP X310 sniffer). Plugged into the laptop are: a BLE dongle (yellow) for sniffing BLE signals and a WiFi adapter (blue) for capturing WiFi packets. Background: an Octoclock-G (green) to ensure precise timing and synchronization across all devices. Foreground: a sample phone/UE (white).}
   
    \label{fig:setup}
    \Description{Experimental setup}
   
\end{figure}

\FloatBarrier

\section{Real Device Measurement}

Figures~\ref{fig:ti_lte_combined}--\ref{fig:ti_wifi_combined} shows the per device
transmission and randomization intervals distribution described in Section~\ref{section:characterization}.

\begin{figure}[H]
 	\includegraphics[width=\columnwidth]{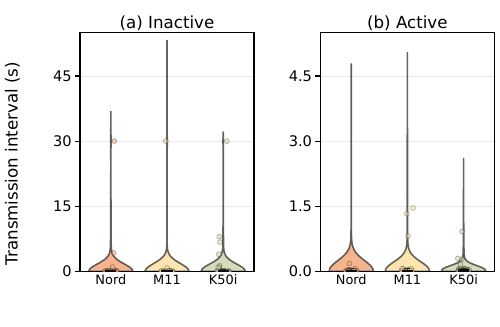}
 	\caption{Transmission Interval: LTE Inactive \& LTE Active} 

 	\label{fig:ti_lte_combined}
    \Description{LTE measurement}
 \end{figure}

\begin{figure}[H]
 	\includegraphics[width=\columnwidth]{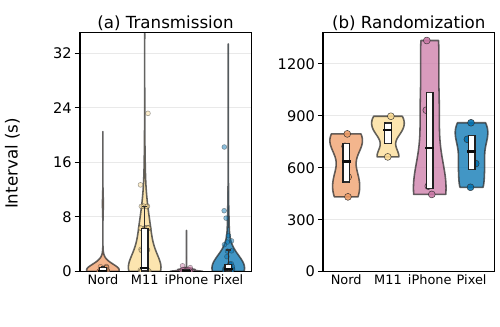}
 	\caption{BLE Transmission \& Randomization Interval}
 	\label{fig:ble_ti_ri_combined}
    \Description{BLE measurement}
 \end{figure}

   \begin{figure*}[tbp]
   	\centering
   	\includegraphics[width=\textwidth]{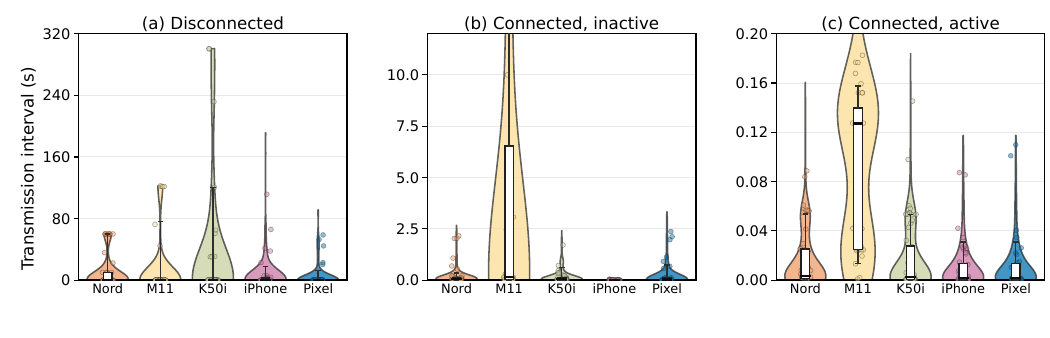}
   	\caption{WiFi Transmission Interval: Disconnected, Connected Inactive, Connected Active}
   	\label{fig:ti_wifi_combined}
    \Description{WiFi measurement}
   \end{figure*}}

\end{document}